\documentclass[aps,prb,twocolumn,showpacs,superscriptaddress]{revtex4-1}
\usepackage{latexsym}
\usepackage{amssymb}
\usepackage{graphicx}
\usepackage{amsmath}
\usepackage{bm}
\usepackage[colorlinks,
          linkcolor=black,
            citecolor=black,
            urlcolor=blue
           ]{hyperref}
\usepackage{verbatim}
\usepackage{slashed}
\usepackage{mathrsfs}
\usepackage{extarrows}
\usepackage{mathtools}
\usepackage{comment}
\usepackage{mathtools,slashed}

\newcommand{\ttheta}[2]{\vartheta\!\!\left[\begin{array}{c}{#1}\\{#2}\end{array}\right]\!\!}

\begin{document}

\title{Bosonization with a background $U(1)$ gauge field}

\author{Yuan Yao}
\author{Yoshiki Fukusumi}
\affiliation{Institute for Solid State Physics, University of Tokyo, Kashiwa, Chiba 277-8581, Japan}

\begin{abstract}

Bosonization is one of the most significant frameworks to analyze fermionic systems.
In this work, we propose a new bosonization of Dirac fermion coupled with $U(1)$ background gauge field.
Our new bosonization is consistent with gauge invariance, global chiral anomaly matching and fermion-boson operator correspondence, either of which is not satisfied by previously developed bosonizations. 
The bilinear Dirac-mass term condensation paradox and its generalized form are resolved by our bosonization.  
This new bosonization approach to interacting systems also correctly captures the conformal characters of a significant class of critical lattice models, such as conformal dimensions and the conformal anomaly of XXZ spin chain with twisted boundary condition. 
Our work clarifies the equivalence of fermionic flux insertion and
bosonic background charge insertion for two-dimensional conformal field theory.

\end{abstract}

\maketitle


\section{Introduction}
Lieb-Schultz-Mattis (LSM) theorem and its higher dimensional analog assign strong constraints on
gapped quantum systems~\cite{LIEB1961407,Oshikawa:2000aa,Hastings:2004aa}.
The key assumption in one of the proofs of these theorems is insensitivity of
the many-body bulk gap under flux insertion~\cite{Oshikawa:2000aa}. 
Moreover such an insensitivity has been recently proven for gapped quantum systems with local Hamiltonians~\cite{PhysRevB.98.155137}. 


On the other hand, renormalization group from conformal fixed point
has been thought to explain the behavior of physical systems \cite{ZAMOLODCHIKOV1989641}.
Furthermore, how critical theories respond to flux insertion process is also of intense interest for LSM-related theorems themselves. For example, LSM theorem and its higher symmetry generalizations can be understood by chiral symmetry anomalies in the presence of nontrivial gauge fields corresponding to flux insertion at the lattice level~\cite{PhysRevB.99.014402, PhysRevB.96.195105, Yao:2018kel}.
Hence it might be natural to guess that such flux insertions or twistings of boundary condition do
not change the behavior of fixed points when we keep in mind the insensitivity of the gapped system.
We will show this naive guess is wrong
and can be explained by bosonization
with background gauge field. 
More precisely for (gapless) critical systems, in the analog of gapped cases, how the spectrum characters, e.g. conformal weight of corresponding conformal field theories, vary upon the twisted boundary condition by flux insertion. That is also essential to several known applications, such as Gaussian fixed point determination for XXZ spin chain~\cite{Fukui:1996aa, Kitazawa:1997aa} and calculating conformal weights~\cite{ALCARAZ1988280}. 
Similar behaviors have already been observed for $6$ and $19$ vertex models which correspond
to one-dimensional (1d) spin chain by quantum statistical correspondence \cite{Klumper:1991aa}.
Within all the applications above, e.g. anomaly manifestation of LSM theorem and XXZ chains with twisted boundary conditions, it is necessary to obtain a critical theory in a background gauge field. 

Historically, the bosonization of Dirac fermion and massive Thirring
model have been widely considered \cite{PhysRevD.11.3026, PhysRevD.11.2088}.
We can think of 1d XXZ Heisenberg model as a realization of
this bosonization \cite{LUKYANOV1997571}. 

In this context, we can think of coset $G/H$ Wess-Zumino-Witten models as the description of bosonic 
coupling of the gauge fields \cite{GAWEDZKI1988119}.
In operator formalism and free field representations, coset construction changes the energy momentum
tensor $T_{G/H}=T_{G}-T_{H}$ \cite{GOODARD198588}. 
In path integral formalism, it can be represented by so-called gauged WZW models and be thought as bosonic theories coupled with the background gauge fields.


However, the equivalence of such a ``conventionally gauged'' WZW model with an action $S_\text{WZW}[A]$ and the adjoint-represented fermion in the corresponding gauge field with $S_\text{Dirac}[A]$ is questioned by Smilga (and N. Nekrasov)~\cite{Smilga:1994aa,Smilga:1996aa}.
It stems from {an apparent contradiction in the} behaviors of the fermionic bilinear condensation between these two models. More specifically speaking, the Dirac mass term does not gain expectation value in the presence of more than two instantons, while {the bosonized term corresponding to the Dirac mass}  is always condensed. Thus the conventionally gauged WZW models {cannot be} bosonization of complex fermions in the presence of general gauge field configurations, e.g. gauge field with nonzero instantons~\cite{Smilga:1994aa,Smilga:1996aa}. 
Moreover, it is shown that
the $U(1)$ boson obtains mass term by gauging out
background gauge field. 
This result itself is quite different from $G/G$ coset WZW model description which
results in topological field theory by gauging out the background gauge field \cite{BLAU1993345, De-Fromont:2014aa}.
Therefore, it is necessary to modify the bosonic gauged WZW models so that they can produce a consistent path integral with their fermionic counterparts. 
Furthermore, as we will see in this paper, the quantum anomaly, e.g. global chiral anomaly of Dirac fermions coupled to background gauge field cannot be reproduced in the gauged WZW model. 
Unfortunately, related to LSM theorem, almost no condensed matter physicist
has ever paid attention to such an inconsistency of bosonization.
However,  the functional bosonization, which results in the same form of the bosonization by Smilga
in some cases, is considered
and applied to some variety of fermionic systems coupled with gauge fields in higher dimensions~\cite{PhysRevB.87.085132}.

\begin{figure}
\begin{center}
\includegraphics[width=6cm,pagebox=cropbox,clip]{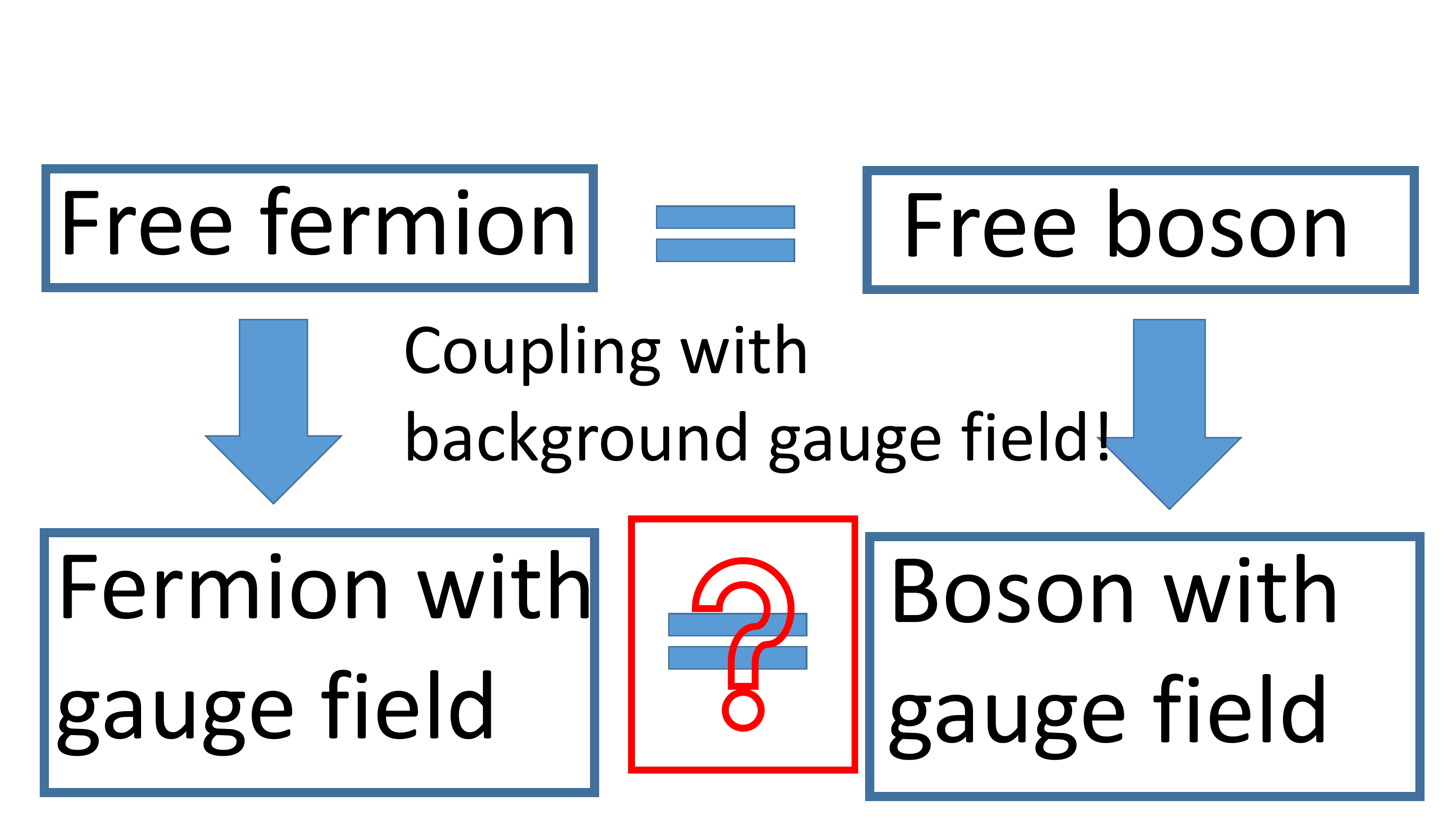}
\caption{Nontriviality of bosonization with background gauge field}\label{fig}
\end{center}
\end{figure}

{A bosonization of Dirac fermion coupled with a background $U(1)$ gauge field with a globally defined $U(1)$ vector potential has been obtained by Fujikawa method~\cite{Fujikawa:2004aa,Fujikawa:2004ab}. 
However, such a condition on gauge fields is generically broken on compact closed manifolds, e.g. gauge fields on a torus with a nontrivial $U(1)$ Chern number (See FIG.~\ref{fig}). }

In this paper, we extend such a bosonization of Dirac fermion coupled with
$U(1)$ gauge field and propose the following new bosonization action on a torus $T^2$ with Euclidean signature, which has two fundamental cycles:
\begin{eqnarray}
S_\text{boson}[A]&=&\int_{T^2}\left[\frac{1}{8\pi}(\partial\varphi)^2-i\frac{e}{2\pi}\varphi dA\right]\nonumber\\
&&-i\frac{\epsilon^{\mu\nu}}{2\pi}\left(\int_{\text{cycle}_\mu}d\varphi\right)\left(\int_{\text{cycle}_\nu}eA\right). 
\end{eqnarray}
We will further see that the contradictions of Dirac-mass condensation mentioned above is resolved and directly related to neutrality conditions in path integral formalism of bosonic model. 
Furthermore, our formalism is superior to that of
Smilga, in the point that we consider the compactification condition of 
free boson on a Riemann surface with a nontrivial genus.
Otherwise, as we will show later, the current-operator correspondence cannot be satisfied between the bosonic and the fermionic sides. 
More precisely, we have considered the contribution of all topological sector of the system, e.g. nonzero winding numbers of field configurations, in the second part of the formula above. 
Although such a winding-type term seems complicated and redundant, it is essential to its topological quantum field theory (TQFT) correspondence in one higher dimension as exposed later. 
In addition, this important term also plays a central role when we generalize the gauge field to be chiral $U(1)_\text{L,R}$ and we will see that our bosonization can reproduce the same $U(1)_\text{L,R}$ charges as other recently developed bosonization transformations~\cite{Thorngren:2018aa}. 

Moreover, our formalism can explain the existing results of the lattice model, XXZ spin chain
with twisted boundary condition.
Unfortunately, our formalism shows functional bosonization and
other known bosonizations are insufficient in the sense that they cannot correctly capture various properties, e.g. partition functions, of
the simplest example, 1d spin chain with twisted boundary condition or
flux insertion.

The organization of the rest of the paper is following.
In Part.~\ref{bosonization}, we motivate and introduce new bosonization
with background gauge field through which we resolve the bilinear-term condensation paradox in its general form.
We apply our formalism to XXZ Heisenberg model
with twisted boundary condition in Part.~\ref{spin chain}.
Then we state conclusions and future direction of
bosonization with background gauge field in Sec.~\ref{conclusions}.

\label{bosonization}
\section{Bosonization with vanishing gauge field}
In this part, the original bosonization without any background $U(1)$ electromagnetic gauge field is briefly reviewed and summarized below to fix the notation for the following discussion. To make the paper self-contained, we first give the Minkowskian Lagrangians: 
\begin{eqnarray}
S_0^\text{(b)}&=&-\int dtdx\frac{1}{8\pi}\left[\partial_\mu\varphi(t,x)\partial^\mu\varphi(t,x)\right]\nonumber\\
S_0^\text{(f)}&=&\int dtdx\,i\psi^\dagger\gamma^0\gamma^\mu\partial_\mu\psi, 
\end{eqnarray}
so that $Z=\int\exp(iS)$, and the Minkowskian signature takes the form as $\eta=\text{diag}(-1,+1)$ with $\{\gamma^\mu,\gamma^\nu\}=-2\eta^{\mu\nu}$ and $\gamma$'s being real, e.g. $\gamma^0=\sigma_1$ and $\gamma^1=i\sigma_2$ where $\vec{\sigma}$ denotes the Pauli matrices. Then the chirality can be defined as $\gamma^3\equiv\gamma^0\gamma^1=-\sigma_3$. 

First, we will normalize several constants and fix the constant conventions. 

When $A_\mu=0$ the external charge $U(1)$ electromagnetic field is vanishing, the duality mapping takes the form as: 
\begin{eqnarray}
\psi(z)&=&\frac{1}{\sqrt{L}}:\!\exp[-i\phi(z)]\!\!:, \\
\bar{\psi}(\bar{z})&=&\frac{1}{\sqrt{L}}:\!\exp[i\bar{\phi}(\bar{z})]\!\!:, \\
\varphi&\equiv&\phi(z)+\bar{\phi}(\bar{z}), \\
\label{compactification}
\varphi&\sim&\varphi+2\pi, 
\end{eqnarray}
in which ``$::$'' denotes normal ordering and $z\equiv x^1+ix^0$ in Euclidean signature, namely $(x^0,x^1)\equiv(it,x)$, and the system scale $L$ is included so that the $\psi(z)$ and $\bar{\psi}(\bar{z})$ have a scaling independent correlation function, where bars denote anti-holomorphism. Eq.~(\ref{compactification}) also normalizes the radius~\cite{Francesco:2012aa} of $\varphi$ to be unit and we will use such a convention throughout this paper. 
\begin{eqnarray}
S_0^\text{(b)}&=&\int\frac{1}{8\pi}(\partial\varphi)^2;\\
S_0^\text{(f)}
&=&\int i\psi^\dagger\gamma^0\left(\gamma^0i\partial_0+\gamma^1\partial_1\right)\psi, 
\end{eqnarray}
where we write $\gamma^\mu$ in its Minkowskian form while space-time coordinates in the Euclidean signature, which is the reason that the form of $S_0^\text{(f)}$ is asymmetric, and $Z=\int\exp(-S)$. 
We can also write down the correspondence of $U(1)$ electromagnetic current operators: 
\begin{eqnarray}
\label{current}
eJ^\nu=\frac{e}{2\pi}\epsilon^{\mu\nu}\partial_\mu\varphi. 
\end{eqnarray}

\begin{figure}
\begin{center}
\includegraphics[width=8cm,pagebox=cropbox,clip]{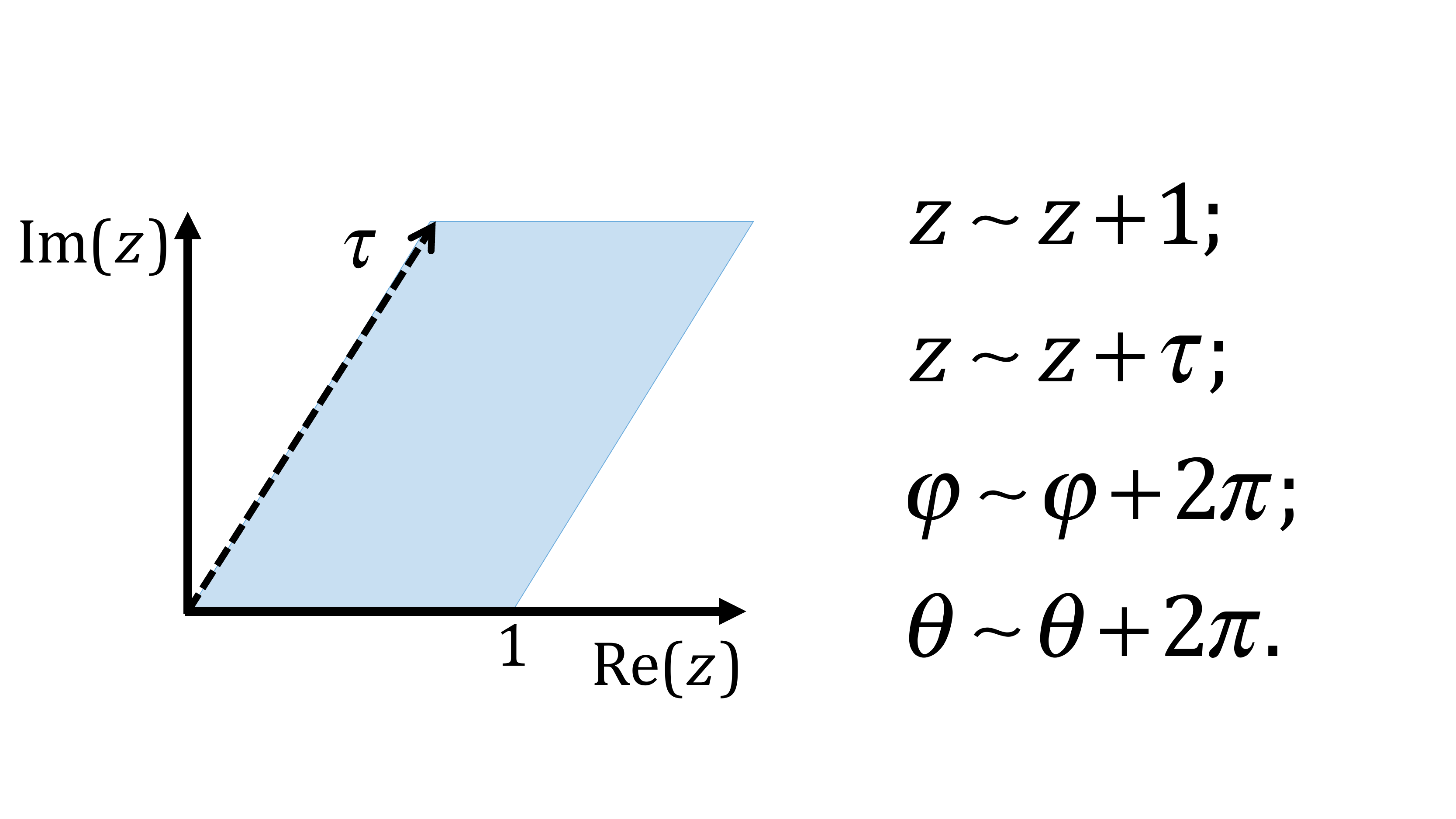}
\caption{Compactification of a boson on a torus parametrized by $\tau$ where $\theta$ is the dual field to $\varphi$ (c.f. Appendix~\ref{TL_liquid}). }
\label{torus}
\end{center}
\end{figure}

Bosonization is completed by showing the two theories are equivalent in the algebraic sense on the torus $T^2$ parametrized by $\tau$ in FIG.~(\ref{torus}), namely possessing the same spectrum. 
To do so, we must sum up all the winding numbers on the bosonic side: 
\begin{eqnarray}
&&Z_0^\text{(b)}=\sum_{n,n'\in\mathbb{Z}}\int\mathscr{D}\varphi\exp(-S_0^\text{(b)}[\varphi]|_{n,n'}),\nonumber\\
&&\varphi(z,\bar{z})=\varphi(z+1,\bar{z}+1)-2\pi n;\nonumber\\
&&\varphi(z,\bar{z})=\varphi(z+\tau,\bar{z}+\bar{\tau})-2\pi n'. 
\end{eqnarray}
It has been proven that the spectrum is equivalent with the fermionic one as long as we sum up the spin structures of Dirac fermion: 
\begin{eqnarray}
&&Z_0^\text{(f)}=\sum_{s_1,s_2\in\{-1,+1\}}\int\mathscr{D}(\psi,\bar{\psi})\exp(-S_0^\text{(f)}[\psi,\bar{\psi}]|_{s_1,s_2}); \nonumber\\
\label{spin_structure}
&&\psi(z+1)=-s_1\psi(z);\,\,\bar{\psi}(\bar{z}+1)=-s_1\bar{\psi}(\bar{z});\nonumber\\
&&\psi(z+\tau)=-s_2\psi(z);\,\,\bar{\psi}(\bar{z}+\bar{\tau})=-s_2\bar{\psi}(\bar{z}). 
\end{eqnarray}
Then
\begin{eqnarray}
\label{free_partition}
&&Z_0^\text{(b)}(\tau)=Z_0^\text{(f)}(\tau)=\sum_{i=1}^4\frac{1}{2}\left(\left|\frac{\theta_i(\tau)}{\eta(\tau)}\right|^2\right), 
\end{eqnarray}
where the Dedekind function is defined as $\eta(\tau)\equiv q^{1/24}\prod_{n=1}^\infty\left(1-q^n\right)$ and $\theta_i(\tau)$'s are the Theta functions~\cite{Francesco:2012aa} with $q\equiv\exp(i2\pi\tau)$ and $\bar{q}=\exp(-i2\pi\bar{\tau})$. It should be noted in advance that, when there is a nonzero background gauge field, the summation weight in Eq.~(\ref{free_partition}) is not equal for every spin structure to be shown later.

\section{Bosonization with background gauge field}
\label{new_bosonization}
For bosonization on torus without background gauge field, it is sufficient to consider bosonization on a Riemann sphere and the calculations can be related to the case on torus by a standard treatment~\cite{Francesco:2012aa}. 
{However, it turns out there exists no straightforward approach to bosonization with background gauge field.}

In this part, we will discuss the bosonization of a single complex Dirac fermion with external background electromagnetic field in the Minkowskian signature: 
\begin{eqnarray}
S_\text{Dirac}[A]&=&\int dtdx\,i\psi^\dagger\gamma^0\gamma^\mu(\partial_\mu-ieA_\mu)\psi, 
\end{eqnarray}
which, in the Euclidean space-time, becomes
\begin{eqnarray}
S_\text{Dirac}[A]=S_0^\text{(f)}+ie\!\!\int\!\!\left[(\psi^\dagger\gamma^0\gamma^0\psi)A_0+(-i\psi^\dagger\gamma^0\gamma^1\psi)A_1\right], \nonumber
\end{eqnarray}
where we have fixed the notation that only $\gamma$ matrices are Minkowskian while all vector fields $A_\mu$ and space-time are Wick-rotated as Euclidean. 
Since the additional term is $i\int eJ^\mu A_\mu$ and observing Eq.~(\ref{current}), one reasonable candidate of the bosonized action is
\begin{eqnarray}
\label{bosonization_wr}
S'[A]=S^{(b)}_\text{0}+i\int\frac{e}{2\pi}A_\nu\epsilon^{\mu\nu}\partial_{\mu}\varphi. 
\end{eqnarray}
Actually this form of action can be thought of as bosonic coupling to background gauge field. More
detailed calculation can be seen in \cite{Chen:2014aa}.
However, $S'[A]$ has the following two problems that 1) $S'[A]$ is gauge-dependent~\cite{Smilga:1996aa} and 2) it has no chiral anomaly factor. 
In other words, we cannot think of this action as ``bosonization''. 

\subsection{$S'[A]$: Gauge non-invariance}
To see the gauge dependence, let us introduce a uniform electromagnetic tensor field: 
\begin{eqnarray}
F_{01}\equiv\epsilon^{\mu\nu}\partial_\mu A_\nu=\frac{2\pi}{|\text{Im}\tau|}, 
\end{eqnarray}
where $|\text{Im}\tau|$ is the area of the spactime torus. To get a local expression of $A_\mu$, we introduce a Dirac-string singularity, e.g. at $x^*$ by some gauge choice. Then
\begin{eqnarray}
&&i\int\frac{e}{2\pi}A_\nu\epsilon^{\mu\nu}\partial_{\mu}\varphi\nonumber\\
&=&i\int\frac{e}{2\pi}\epsilon^{\mu\nu}\left[\partial_{\mu}(\varphi A_\nu)-(\partial_\mu A_\nu)\varphi\right]\nonumber\\
&=&{i}e[\varphi(x^*)-\varphi_\text{ave}], 
\end{eqnarray}
where $\varphi_\text{ave}\equiv\int\varphi/|\text{Im}\tau|$ is the average value of $\varphi$ upon the torus. We could see that the naive imposing the duality mapping of current in Eq.~(\ref{current}) does not give a gauge invariant theory on the bosonic side. 

\subsection{$S'[A]$: {Mismatch} of the anomaly and partition function}
Furthermore, the action $S'[A]$ suffers from the chiral anomaly enjoyed by the fermionic action. The chiral transformation for the bosonic field is $\varphi\rightarrow\varphi+\text{const.}$. Obviously, the bosonic theory and the partition function defined by $S'[A]$ is invariant under such a transformation. Therefore, its chiral anomaly does not match that of a single complex fermion. One direct result from such an anomaly mismatching is the discrepancy between the partition function obtained from integrating out $\mathscr{D}\varphi$ with $S'[A]$ and the fermionic partition function in the appearance of nonzero flux: $\int eF_{01}/(2\pi)\neq0$. By Atiyah-Singer index theorem, there must exist at least one zero mode of the Dirac operator. Then, formally, 
\begin{eqnarray}
Z_\text{Dirac}[A]\propto\prod_{k\in K}\lambda_k=0, 
\end{eqnarray}
where $\{\lambda_k\}_{k\in K}$ is the spectrum of Dirac operator. 

However, $\int\mathscr{D}\varphi\exp(-S'[A])\neq0$ generically. As a typical example, let us choose the following gauge-field configuration $\{\tilde{A}_\mu\}$: 
\begin{eqnarray}
\left\{\begin{array}{ll}e\tilde{A}^\text{I}_0=0,\,e\tilde{A}^\text{I}_1=\frac{2\pi}{\epsilon}(x_0-\tilde{x}_0),&\text{if }\vec{x}\in U_\text{I};\\
e\tilde{A}^\text{II}_0=e\tilde{A}^\text{II}_1=0, &\text{if }\vec{x}\in U_\text{II}. \end{array}\right.
\end{eqnarray}
Here $U_\text{I}\equiv[\tilde{x}_0,\tilde{x}_0+\epsilon]\times[0,L_1)$ and $U_\text{II}=\bar{U}_\text{I}$ is its complement. It can be calculated that $\int\tilde{F}_{01}/(2\pi)=1$ thereby $Z^\text{(f)}[\tilde{A}]=0$. 

When we take $\epsilon\rightarrow0^+$, it is straightforward to check that $Z'[\tilde{A}]\equiv\int\mathscr{D}\varphi\exp(-S'[\tilde{A}])=\int\mathscr{D}\varphi\exp(-S_0^{(b)})=Z_0^{(b)}$,
where $Z_0^{(b)}$ is the partition function of free boson without background gauge field in Eq.~(\ref{free_partition}) which is nonzero. Therefore, $Z'[\tilde{A}]\neq Z^\text{(f)}[\tilde{A}]$. 

Thus we come to the second candidate of bosonization by a total derivative addition as a counterterm: 
\begin{eqnarray}
S''[A]&=&S'[A]\!-\!i\int\frac{e}{2\pi}\epsilon^{\mu\nu}\partial_\mu[\varphi A_\nu]=S_0^\text{(b)}-i\int\frac{e}{2\pi}\varphi F_{01}, \nonumber
\end{eqnarray}
which is explicitly gauge-invariant since the curvature tensor $F_{01}$ is gauge-independent. 
This form of action was first introduced by~\cite{Faddeev:1976aa} and extensively considered by~\cite{Smilga:1994aa}. It is valid if we think about the trivial topological sector or zero winding number of $\varphi$ and consistent with functional bosonization.

\subsection{$S''[A]$: Mismatching of $U(1)$ electromagnetic current}
However, since $\varphi$ is not single-valued, the current $J^\mu$ might not be properly coupled with $A_\mu$ in $S''[A]$. Indeed, let us take the functional derivative: 
\begin{eqnarray}
\delta_{ieA_\rho}S''[A]&=&\frac{1}{2\pi}\epsilon^{\mu\rho}\partial_\mu\varphi-\delta_{A_\rho}\int\frac{1}{2\pi}\epsilon^{\mu\nu}\partial_\mu[\varphi A_\nu]\nonumber\\
&=&J^\rho+n'\delta(x_0+0^+)\epsilon^{0\rho}+n\delta(x_1+0^+)\epsilon^{1\rho}, \nonumber
\end{eqnarray}
where, without loss of generality, just for simplicity, we have assumed the (Euclidean) rectangular spacetime (before quotiented to the torus) $[0,L_0]\times[0,L_1]$. The form of $\delta$-functions depends how we distribute the unity between two equivalent boundary point and, by no means, will affect the following results. To see why the term ``$-\delta_{A_\rho}\int\epsilon^{\mu\nu}\partial_\mu[\varphi A_\nu]/2\pi$'' only gives the additional boundary current ``$n'\delta(x_0-L^-_0)\epsilon^{0\rho}+n\delta(x_1-L^-_1)\epsilon^{1\rho}$'', we perform the integration in the following form: 
\begin{eqnarray}
\label{total_derivative}
\int_M\frac{1}{2\pi}\epsilon^{\mu\nu}\partial_\mu[\varphi A_\nu]&=&\int_M\frac{1}{2\pi}d(\varphi A)\\
&=&\int_{\partial M}\frac{1}{2\pi}\varphi A+\sum_{i}\int_{\partial U_i}\frac{1}{2\pi}\varphi A_{(i)}. \nonumber
\end{eqnarray}
where we take $M$ as a rectangular from which the torus is made by conventional pasting procedure. We can see that the remaining bulk part ``$\sum_{i}\int_{\partial U_i}\frac{1}{2\pi}\varphi A_{(i)}$'' is gauge dependent, in which $U_i$'s, where $A_{(i)}$ is locally well-defined depending on gauge choices, cover $M$. Then, 
\begin{eqnarray}
\label{bulk_current}
\sum_{i}\int_{\partial U_i}\frac{1}{2\pi}\varphi A_{(i)}&=&\sum_{i,j}\,\!'\int_{\partial U_i\cap\partial U_j}\frac{1}{2\pi}\varphi\,t_{ij}^{-1}dt_{ij}, 
\end{eqnarray}
where $\sum'_{i,j}$ denote no double-counting with proper orientations of $\partial U_i\cap\partial U_j$'s, and $t_{ij}$ is the transition function defined by $A_i=A_j+t_{ij}^{-1}dt_{ij}$. On the other hand, 
$\delta_{A_\rho}(t_{ij}^{-1}dt_{ij})=\delta_{A_\rho}(A_i-A_j)=0$, which implies such a \emph{gauge-dependent} bulk contribution induced by Eq.~(\ref{bulk_current}) vanishes: $\delta_{A_\rho}\left[\sum_{i}\int_{\partial U_i}\varphi A_{(i)}/(2\pi)\right]=0$. The first term $\int_{\partial M}\varphi A/2$ in the last line of Eq.~(\ref{total_derivative}) gives the ``boundary'' current: 
\begin{eqnarray}
-\delta_{A_\rho}\int_{\partial M}\frac{1}{2\pi}\varphi A
&=&\delta_{A_\rho}\left[\int_{z=0}^{z=1}n'A-\int_{z=0}^{z=\tau}nA\right]\nonumber\\
&=&n'\delta(x_0+0^+)\epsilon^{0\rho}+n\delta(x_1+0^+)\epsilon^{1\rho}. \nonumber
\end{eqnarray}

\subsection{$S_\text{boson}[A]$: Cancellation of boundary current}
To cancel the additional ``boundary'' coupling which induces the boundary current $\left(-n'\delta(x_0+0^+)\epsilon^{0\rho}-n\delta(x_1+0^+)\epsilon^{1\rho}\right)$, we \emph{tentatively} take into consideration the following modified action $S_\text{boson}$ so that $(1/i)\delta_{A_\rho}S_\text{boson}=J^\rho$: 
\begin{eqnarray}
&&S_\text{boson}|_{n,n'}=S''[A]+i\!\!\int\!\! eA_\rho\left[\delta_{A_\rho}\int_{\partial M}\frac{1}{2\pi}\varphi A\right]\nonumber\\
\label{boson_1}
&=&\!\!\int\!\left[\frac{1}{8\pi}(\partial\varphi)^2\!-\!i\frac{e}{2\pi}\varphi\epsilon^{\mu\nu}\partial_\mu A_\nu\right]\!\!+\!i2\pi(\!-n'\alpha\!+\!n\beta), 
\end{eqnarray}
where 
\begin{eqnarray}
\label{flat}
\alpha&\equiv&\frac{e}{2\pi}\int_{\text{cycle}_1} dx^1 A_1(x^0=0,x^1);\nonumber\\
\beta&\equiv&\frac{e}{2\pi}\int_{\text{cycle}_0} dx^0 A_0(x^0,x^1=0), 
\end{eqnarray}
where ``$\text{cycle}_{0,1}$'' are two generating cycles of the underlying torus along real axis and $\tau$ direction, respectively. Alternatively in a compact way, 
\begin{eqnarray}
\label{boson_2}
S_\text{boson}[A]&=&S_0^\text{(b)}-i\frac{e}{2\pi}\int_{T^2}\varphi dA\nonumber\\
&&-i\frac{\epsilon^{\mu\nu}}{2\pi}\left(\int_{\text{cycle}_\mu}d\varphi\right)\left(\int_{\text{cycle}_\nu}eA\right). 
\end{eqnarray}
{This is the new action of the bosonized theory we propose in this paper.}

\subsection{$S_\text{boson}[A]$: Gauge invariance, equation of motion and LSM-type arguments}
As one of the several necessary checks, $S_\text{boson}[A]$ obviously still has the correct chiral anomaly factor $\text{exp}(i \nu \int eF_{01})$ as $S''[A]$, with the chiral transformation $\varphi \rightarrow \varphi +2\pi \nu$. In addition, it is also gauge-invariant since the coefficient of $2\pi\alpha$ and $2\pi\beta$ is integer despite of the fact that $\alpha$ and $\beta$ are only gauge invariant modulo $\mathbb{Z}$ or only $(\alpha\!\mod\mathbb{Z})$ and $(\beta\!\mod\mathbb{Z})$ are gauge invariant. Furthermore, $S_\text{boson}[A]$ gives a correct equation of motion:
$\delta_{\varphi}S_\text{boson}[A]=-\partial^2\varphi/(4\pi)-ieF_{01}/(2\pi)=0$ beause
$n'$ and $n$ are integer-valued which implies they are insensitive and invariant for any infinitesimal variation: $\delta_\varphi n'=\delta_\varphi n=0$. Thus, the equation of motion $\partial^2\varphi/(4\pi)=-ieF_{01}/{(2\pi)}$ is exactly the equation of motion of axial current on the fermionic side and the appearance of ``$i$'' on the right-hand side is due to the Wick rotation of $A_\mu$. 

{It is consistent with the anomaly argument of LSM theorem~\cite{PhysRevB.96.195105,Yao:2018kel}. 
Namely, the lattice translational transformation of the underlying tight-binding electronic system at low-energy limit $\mathbb{Z}_\text{trans}:\varphi\rightarrow\varphi+2\pi\nu$ with $\nu$ the particle number per unit cell~\cite{PhysRevB.96.195105} indeed induces a phase change of the partition function $Z_\text{boson}[A]\equiv\int\mathscr{D}\varphi\exp(-S_\text{boson}[A])$ by $\exp\left(i\nu\int eF_{01}\right)$ characterizing the 't Hooft anomaly between $U(1)$ and $\mathbb{Z}_\text{trans}$. Such an anomaly can be directly seen by Eq.~(\ref{boson_2}) due to the $\varphi$-linear term proportional to $\int\varphi dA$. This phase term is trivial for arbitrary $U(1)$ gauge field configuration if and only if $\nu\in\mathbb{Z}$ or the particle per unit cell is an integer, due to the quantization $\int eF_{01}/2\pi\in\mathbb{Z}$ on torus. Assuming that a non-trivial anomaly phase implies that the system cannot be symmetrically gapped with a unique ground state~\cite{PhysRevB.96.195105,Yao:2018kel}, we reproduces the LSM theorem that the non-integer filling fraction does not permitted a symmetric insulating phase if $U(1)$ and $\mathbb{Z}_\text{trans}$ are respected~\cite{LIEB1961407,Oshikawa:2000aa,Hastings:2004aa}. Alternatively, by a $2\pi$-flux insertion procedure, we have also obtained a lattice momentum change $2\pi\nu/L\mod2\pi/L$ by Eq.~(\ref{momentum_change}). We also obtain the LSM theorem by the flux insertion argument~\cite{Oshikawa:2000aa}. }

\subsection{TQFT/CFT interpretation of $S_\text{boson}[A]$}
The free bosonic model or its generalization Wess-Zumino-Witten (WZW) model can be reduced from Chern-Simons (CS) theory, which provides a TQFT/CFT interpretation on both theories defined in different dimensions. It implies that we can obtain a WZW model as the boundary theory of a CS theory~\cite{Witten:1989aa,moore:1989,Nayak:2008aa}. However, such a dimension-reduction derivation, when space-time manifold is taken as $T^2$, is subtle because of the typical ``half-lives, half-dies'' principle in three-manifold topology. Namely, it is stated that one ``$\mathbb{Z}$'' component of the first homological group $H_1(T^2,\mathbb{Z})\cong\mathbb{Z}\times\mathbb{Z}$ of $T^2$ will be trivialized when $T^2$ is extended to some three-dimensional compact manifold $M$ such that $\partial M=T^2$. For example, the nonzero winding of $\varphi(x)$ along the cycle which is trivialized by the dimension extension obstructs the extension of field configuration $\varphi(x)$ onto $M$. Conversely, the corresponding dimension reduction from a CS theory on $M$ cannot produce the whole field configuration of the WZW model defined solely on $T^2$. 

Although the TQFT/CFT correspondence above is problematic when we extend the manifold from $T^2$ to $M$ in the foregoing sense, one could still see the reasonable addition of the ``strange'' second term $-i{\epsilon^{\mu\nu}}/{2\pi}\left(\int_{\text{cycle}_\mu}d\varphi\right)\left(\int_{\text{cycle}_\nu}eA\right)$ of Eq.~(\ref{boson_2}). 

To avoid the ``half-lives, half-dies'' principle, let us consider an infinitely long torus where $\tau_2\equiv\text{Im}\tau\rightarrow+\infty$ so that the nontrivial winding mode of $\varphi(x)$ in the spatial direction $x_1$ will be gapped thereby vanishing in the resultant partition function of CFT on $T^2$. It implies it is a reasonable approximation to treat the torus $T^2$ as the boundary of toroid $M\equiv D^2\times S^1$ of which the infinitely-long ``$S^1$'' parametrizes time. We further assume that $A$ can be extended to the bulk $M$ where the $U(1)$ monopoles, which create integral net fluxes on $T^2$, are defined conventionally by point defects within $M$. 
We can obtain, by the Bianchi identity $ddA=0$ in $M$ and omitting the spatial winding of $\varphi(x)$ in Eq.~(\ref{boson_2}), 
\begin{eqnarray}
\label{boson_3}
S_\text{boson}[A]
=S_0^\text{(b)}-i\frac{e}{2\pi}\int_{M}d\varphi\wedge dA, 
\end{eqnarray}
where we can see that the complicated term is essential to have the compact form above of higher dimensional extension. Furthermore, if we require the extended background field $A$ on $M$ to satisfy $F_{rz}\equiv\partial_r A_z-\partial_zA_r=0$ in which $r$ is the radial coordinate of $D^2$ and $z=x^1+ix^0$ as before and impose the Neumann boundary condition $\partial_r\varphi|_{\partial M}=0$, we can obtain the following holographic interpretation of our bosonization scheme: 
\begin{eqnarray}
&&\int\mathscr{D}\varphi\exp\left(-S_\text{boson}[A]\right)\nonumber\\
&=&\int\mathscr{D}\varphi\exp\left[-\int_{D^2\times S^1} drdzd\bar{z}2i\partial_r\left(\partial_z\varphi\partial_{\bar{z}}\varphi\right)/(8\pi)\right.\nonumber\\
&&\left.+ie\int_{D^2\times S^1}drdzd\bar{z}\left(\partial_r\varphi F_{z\bar{z}}-\partial_z\varphi F_{r\bar{z}}\right)\right]/(2\pi)\nonumber\\
&=&\int\mathscr{D}B_r\mathscr{D}B_z\cdot\delta(F_{B,rz})\exp\left[\frac{1}{4\pi i}\int_{D^2\times S^1}drdzd\bar{z}\epsilon_{(rz)}^{ij}\right.\nonumber\\
&&\left.B_i\partial_{\bar{z}}B_j-\frac{e}{2\pi}\int_{D^2\times S^1}B\wedge dA\right]\nonumber\\
&=&\int\mathscr{D}B_\mu\exp\left[-\int_M\left(\frac{1}{4\pi}B\wedge dB+\frac{e}{2\pi}B\wedge dA\right)\right], 
\end{eqnarray}
where $B_\mu=(B_r,B_z,B_{\bar{z}})$ and the superscripts of antisymmetric tensor $\epsilon^{ij}_{(rz)}$ take values in $\{r,z\}$ with $\epsilon^{rz}_{(rz)}=-\epsilon^{zr}_{(rz)}=1$, and $F_{B,rz}\equiv\partial_rB_z-\partial_zB_r$ with $B_{\bar{z}}$ serving as the Lagrange multiplier of $\delta(F_{B,rz})$. 
It should be noted that here the ``temporal'' coordinate $r$ is set to be the extra dimension and, if we set the spatial ones, e.g. $x^1$, to be radial one instead, the corresponding bosonic theory will be chiral~\cite{Shifman:1999aa}. 

Thus, it implies Chern-Simons-BF (CSBF) theory~\cite{Ikeda:2003aa,Seiberg:2016aa} is the dual TQFT of our bosonization: 
\begin{eqnarray}
\label{boson_4}
S_\text{boson}[A]&\leftrightarrow&\int\frac{1}{4\pi}B\wedge dB+\frac{e}{2\pi}B\wedge dA\equiv S_\text{CS-BF}[B,A], \nonumber 
\end{eqnarray}
with the duality mapping $-id\varphi\leftrightarrow B$ satisfying $B_r|_{\partial M}=0$ or equivalently Neumann boundary condition $\partial_r\varphi|_{\partial M}=0$ and the extension requirement that $F_{rz}=0$ for the background charge $U(1)$ gauge field $A$. 
\section{Spectral equivalence}
\subsection{Spectrum with a flat background gauge field}
We will take a simple case so that the duality could be seen readily. The background gauge field will be taken flat so that $F_{01}=0$. For later convenience, let us take a more general Luttinger parameter as $1/8\pi\rightarrow1/8\pi K$ or $S_0^\text{(b)}\rightarrow S^0_\text{T-L}\equiv S_0^\text{(b)}/K$, though the current interest is $K=1$. Such a generalized model corresponds to a type of interacting fermions and we will later show that the remaining terms of $S_\text{boson}[A]$ indeed do not gain renormalization by $K$ due to topological reasons (c.f. Eq.~(\ref{non-re})).  

We can calculate the partition function associated with $S_\text{boson}[K;\alpha,\beta]$ as (c.f. Eq.~\ref{flatpartition_1})
\begin{eqnarray}
\label{flatpartition}
&&Z_\text{boson}[K,A_\text{flat}]\\
&=&\sum_{n,k}\exp(-i2\pi n\beta)q^{\frac{K}{2}\left(k+\frac{n}{2K}+\alpha\right)^2}\bar{q}^{\frac{K}{2}\left(k-\frac{n}{2K}+\alpha\right)^2}/|\eta(\tau)|^2. \nonumber
\end{eqnarray}
We can identify the charge of $\mathbb{Z}_2$ symmetry with a certain operator labelled by $(n,k)$ as
\begin{eqnarray}
\label{z2_charge}
Q_{\mathbb{Z}_2}=n, 
\end{eqnarray}
due to $\beta=1/2$ giving $\exp(-i2\pi n\beta)=(-1)^n$ which is its fermion number parity. 
This $\mathbb{Z}_{2}$ symmetry is generated by $(-1)^{Q_{\mathbb{Z}_2}}:\Psi\rightarrow-\Psi$ the fermion number parity transformation or $\Theta\rightarrow\Theta+\pi$ bosonically.

Then we arrive at the following result at the free fermion point $K=1$: 
\begin{eqnarray}
\label{flat_duality}
&&\!Z_\text{boson}[A_\text{flat}]\!\!=\!\!\!\!\!\!\sum_{f_{0,1}\in\{0,1/2\}}\!\!\!\!\!\!\frac{(-1)^{\delta_{f_0+f_1,1}}}{2|\eta(\tau)|^2}\!\!\left\{\left|{\ttheta{\alpha+f_1}{-(\beta+f_2)}(\tau)}\right|^2\!\right\}\nonumber\\
&=&\frac{1}{2}\left\{Z^{+,+}_\text{Dirac}+Z^{+,-}_\text{Dirac}+Z^{-,+}_\text{Dirac}-Z^{-,-}_\text{Dirac}\right\}[\alpha,\beta], 
\end{eqnarray}
where $Z^{s_1,s_2}_\text{Dirac}$ labels the the Dirac partition function~\cite{Alvarez-Gaume:1986aa} with the spin structure $(s_1,s_2)$ defined in Eq.~(\ref{spin_structure}) and
$\ttheta{\alpha}{-\beta}(\tau)\equiv\sum_{n\in\mathbb{Z}}\exp\left[i\pi(n+\alpha)^2\tau-i2\pi\beta\right]$ is the generalized Theta function~\cite{Blumenhagen:2012aa}. 
To obtain the fermionization or the inverse of the bosonization which completes the bosonization procedure, we can made use of the $\mathbb{Z}_2$ transformation defined above, whose charge is obtained in Eq.~(\ref{z2_charge}). We can apply this $Z_2$ transformation onto the Hilbert space as an operator, equivalent to inserting $(-1)^{Q_{\mathbb{Z}_2}}=\exp\left(-{i}\int_{\text{cycle}_1}d\varphi/2\right)$
into the bosonic path integral. Similarly, we can also apply this $\mathbb{Z}_2$ transformation to twist the bosonic wave function spatially by a defect line operator, equivalently inserting 
$I_{\mathbb{Z}_2}\equiv\exp\left({i}\int_{\text{cycle}_0}d\varphi/2\right)$
into the bosonic path integral. 

Then we can label the corresponding $\mathbb{Z}_2$ sectors by $\mathbb{Z}_\text{boson}^{w_1,w_2}$ where $w_1\&w_2\in\{\pm\}$ with ``$+$'' no $\mathbb{Z}_2$ twisting whereas ``$-$'' a $\mathbb{Z}_2$ twisting, denotes whether the $\mathbb{Z}_2$ generator is operated spatially and temporally, respectively. Specially, $Z_\text{boson}[A_\text{flat}]$ in Eq.~(\ref{flat_duality}) is $Z_\text{boson}^{+,+}$. 
Therefore, 
\begin{eqnarray}
\label{matrix_bosonization}
\left(\begin{array}{c}Z_\text{boson}^{+,+}\\Z_\text{boson}^{+,-}\\Z_\text{boson}^{-,+}\\Z_\text{boson}^{-,-}\end{array}\right)=\frac{1}{2}\left(\begin{array}{cccc}1&1&1&-1\\1&1&-1&1\\1&-1&1&1\\-1&1&1&1\end{array}\right)\left(\begin{array}{c}Z_\text{Dirac}^{+,+}\\Z_\text{Dirac}^{+,-}\\Z_\text{Dirac}^{-,+}\\Z_\text{Dirac}^{-,-}\end{array}\right) 
\end{eqnarray}
for the flat background gauge field and we can defined a matrix $W_{(w_1,w_2),(s_1,s_2)}$ so that
\begin{eqnarray}
Z_\text{boson}^{w_1,w_2}[\alpha,\beta]=\sum_{(s'_1,s'_2)}W_{(w_1,w_2),(s'_1,s'_2)}Z_\text{Dirac}^{s_1',s_2'}[\alpha,\beta], 
\end{eqnarray}
and similarly, the fermionization takes the form as
\begin{eqnarray}
Z_\text{Dirac}^{s_1,s_2}[\alpha,\beta]=\sum_{(w'_1,w'_2)}W^{-1}_{(s_1,s_2),(w'_1,w'_2)}Z_\text{boson}^{w_1',w_2'}[\alpha,\beta], 
\end{eqnarray}
where numerically $W=W^{-1}$. 
{Such an invertibility of $W$ exactly confirms the conjecture that weighted sum of fermion partition functions over various spin structure being the boson partition function in Eq.~(\ref{flat_duality}) implies the bosonization of a fermion partition with a single spin structure (specified by $s_{1,2}$, $\alpha$ and $\beta$ on torus) in the earlier discussions on twisted fermions~\cite{Alvarez-Gaume:1986aa,Stone:1994aa}. }
The bosonization and fermionization above exactly reproduce the same results with $\alpha=\beta=0$ except for the last column of $W$ matrix which does not matter with a vanishing background gauge field~\cite{Gaiotto:2015aa,Gaiotto:2016aa,Kapustin:2017aa,Thorngren:2018aa}. 

\subsection{Bosonization: Duality of partition function}
An observation of the flat-connection case in Eq.~(\ref{flat_duality}) implies the bosonic partition function cannot be dualized to some fermionic one unless all the possible spin structures are summed up by a weight determined by matrix $W_{(w_1,w_2),(s_1,s_2)}$ and its inverse. 
Hence, assuming these weights only depend on spin structures, we could propose that, for general fluctuating $\{A_\mu\}$ gauge field configurations, 
\begin{eqnarray}
&&Z_\text{Dirac}^{s_1,s_2}[A]=\int\mathscr{D}(\psi,\bar{\psi})\exp(-S_\text{Dirac}[\psi,\bar{\psi},A]|_{s_1,s_2}); \nonumber\\
&&\psi(z+1)=-s_1\psi(z);\,\,\bar{\psi}(\bar{z}+1)=-s_1\bar{\psi}(\bar{z});\nonumber\\
&&\psi(z+\tau)=-s_2\psi(z);\,\,\bar{\psi}(\bar{z}+\bar{\tau})=-s_2\bar{\psi}(\bar{z})
\end{eqnarray}
and its dual
\begin{eqnarray}
&&Z_\text{boson}^{w_1,w_2}[A]=\sum_{n,n'\in\mathbb{Z}}\int_{w_1,w_2}\mathscr{D}\varphi\exp(-S_\text{boson}[\varphi,A]|_{n,n'}),\nonumber\\
&&\varphi(z,\bar{z})=\varphi(z+1,\bar{z}+1)-2\pi n;\nonumber\\
&&\varphi(z,\bar{z})=\varphi(z+\tau,\bar{z}+\bar{\tau})-2\pi n', 
\end{eqnarray}
are related by matrix $W$ defined by Eq.~(\ref{matrix_bosonization}): 
\begin{eqnarray}
Z_\text{boson}^{w_1,w_2}[A]&=&\sum_{(s'_1,s'_2)}W_{(w_1,w_2),(s'_1,s'_2)}Z_\text{Dirac}^{s_1',s_2'}[A]; \nonumber\\
Z_\text{Dirac}^{s_1,s_2}[A]&=&\sum_{(w'_1,w'_2)}W^{-1}_{(s_1,s_2),(w'_1,w'_2)}Z_\text{boson}^{w_1',w_2'}[A], 
\end{eqnarray}
which can be generalized to the interacting fermion with a general $K$ not necessarily $1$. 

Let us furthermore gauge the bosonic $\mathbb{Z}_2$ symmetry or equivalently impose the identification $\theta\sim\theta+\pi$. Then
\begin{eqnarray}
\label{z2_gauged}
Z^{\mathbb{Z}_2\text{-gauge}}_\text{boson}[A]\!&=&\!\frac{1}{2}\sum_{w_{1,2}}Z_\text{boson}^{w_1,w_2}[A]=\frac{1}{2}\sum_{s_{1,2}}Z_\text{Dirac}^{s_1,s_2}[A], 
\end{eqnarray}
which is exactly the Dirac fermion gauged by $\mathbb{Z}_2$ fermion number parity. 


\subsection{Chiral $U(1)_\text{L,R}$ symmetries and their extensions after bosonized}

Let us consider a natural generalization of the background gauge field that is enlarged to the chiral $U(1)_\text{L}\times U(1)_\text{R}$. In the following, we will investigate how such a (fermionic) chiral symmetry $U(1)_\text{L,R}$ is represented bosonically and we will show our bosonization can reproduce exactly the same result as other bosonization approaches~\cite{Thorngren:2018aa}. The results within this part are applicable to general $K$ not necessarily the free fermion point $K=1$. We take the following convention for the chiral gauge coupling to the free fermion part as:
\begin{eqnarray}
S_{0\text{chiral}}^\text{(f)}&=&\int dtdx\,i\psi^\dagger\gamma^0\gamma^\mu\left(\partial_\mu-ieP_\text{L}A_{\text{L}\mu}-ieP_\text{R}A_{\text{R}\mu}\right)\psi, \nonumber
\end{eqnarray}
where $U(1)_\text{L,R}$ connections $A_\text{L}$ and $A_\text{R}$ are not necessarily equal and $P_\text{L,R}\equiv(1\mp\gamma^3)/2$. 
It is straightforward to obtain the corresponding action after one notice the correspondences $\psi(z)\propto:~\!\!\exp[-i(\varphi+2\theta)/2]~\!\!:$ with $\bar{\psi}(\bar{z})\propto:~\!\!\exp[i(\varphi-2\theta)/2]~\!\!:$ where we still take a general Luttinger parameter tuned by $K$ with $\theta$ compactified by $2\pi$. Then
\begin{widetext}
\begin{eqnarray}
\label{boson_chiral}
&&S_\text{boson}[K,A_\text{L,R}]=S^0_\text{T-L}-i\frac{e}{2\pi}\int_{T^2}\left[\frac{1}{2}\left(\varphi+2\theta\right)dA_\text{R}+\frac{1}{2}\left(\varphi-2\theta\right)dA_\text{L}\right]\nonumber\\
&&-i\frac{\epsilon^{\mu\nu}}{4\pi}\left\{\left[\int_{\text{cycle}_\mu}d\left(\varphi+2\theta\right)\right]\left(\int_{\text{cycle}_\nu}eA_\text{R}\right)+\left[\int_{\text{cycle}_\mu}d\left(\varphi-2\theta\right)\right]\left(\int_{\text{cycle}_\nu}eA_\text{L}\right)\right\}. 
\end{eqnarray}
\end{widetext}
Then we take $A_\text{L}=0$ and $dA_\text{R}=0$ with $\int_{\text{cycle}_\nu}eA_\text{R}=2\pi\delta_{\nu,0}\beta$ to obtain the $U(1)_\text{R}$ chiral charge. Hence, 
\begin{eqnarray}
\label{boson_chiral}
&&S_\text{boson,R}[K,\beta]=S^0_\text{T-L}+i\frac{\beta}{2}\int_{\text{cycle}_1}d\left(\varphi+2\theta\right). 
\end{eqnarray}
By the use of 
$\partial_{\bar{z}}(\varphi+2K\theta)=\partial_z(\varphi-2K\theta)=0$~\cite{Suppl,Francesco:2012aa}, we can arrive at that the additional term proportional $i\beta/2\cdot\int_{\text{cycle}_1}d(2\theta)$ can be cancelled by the following changing of variable, up till a real constant in the action which will be set to zero required by unchanged central charge due to $\alpha=0$, 
$\varphi\rightarrow\tilde{\varphi}=\varphi+2\pi \beta\frac{\bar{z}-z}{\bar{\tau}-\tau}$, 
or, equivalently~\cite{Suppl}, $n'\rightarrow\tilde{n}'=n'+\beta$. Then we obtain by $q\equiv\exp(i2\pi\tau)$ and $\bar{q}\equiv\exp(-i2\pi\bar{\tau})$, (c.f. Appendix~\ref{z_flat})
\begin{eqnarray}
\label{flatpartition}
Z_\text{boson,R}[K,\beta]&=&\frac{1}{|\eta(\tau)|^2}\sum_{n,k}\exp\left[-i2\pi(n/2-k)\beta\right]\nonumber\\
&&\cdot q^{\frac{K}{2}\left(k+\frac{n}{2K}\right)^2}\bar{q}^{\frac{K}{2}\left(k-\frac{n}{2K}\right)^2}, 
\end{eqnarray}
from which we can directly read off the $U(1)_\text{R}$ charge, and similar calculations yield $U(1)_\text{L}$ charge: 
\begin{eqnarray}
Q_\text{R}=n/2-k;\,\,Q_\text{L}=n/2+k, 
\end{eqnarray}
which are exactly the same as the results in the point of view of quantum anomaly~\cite{Thorngren:2018aa}. 

For general bosonic $\mathbb{Z}_2$-sectors $Z_\text{boson}^{w_1,w_2}$ to be inserted by $\exp(i2\pi Q_\text{L,R})$ in the path integral, the effects of such insertions are: 
\begin{eqnarray}
&&Z_\text{boson}^{w_1,w_2}[\alpha,\beta]\rightarrow Z_\text{boson}^{w_1,-w_2}[\alpha,\beta], 
\end{eqnarray}
which implies that
\begin{eqnarray}
\label{gauge_gauge}
Z^{\mathbb{Z}_2\text{-gauge}}_\text{boson}[\alpha,\beta]&\rightarrow& Z^{\mathbb{Z}_2\text{-gauge}}_\text{boson}[\alpha,\beta], \\
\label{charge_relation}
\exp(i2\pi Q_\text{L,R})&=&(-1)^{Q_{\mathbb{Z}_2}}, 
\end{eqnarray}
and the $\mathbb{Z}_2$ gauged bosonic theory has a bosonic spectrum that is natural in the viewpoint of Eq.~(\ref{z2_gauged}). 
It also means that the fermionic $U(1)_\text{L,R}$ is extended by $\mathbb{Z}_2$ on the bosonic side and we expect such a relation is also held when the gauge field $\{A_\mu\}$ is fluctuating. Due to such a symmetry extension, a bosonized theory would have less symmetry anomalies than its fermionic partner. Thus, together with the equivalence in Eq.~(\ref{gauge_gauge}), it implies there exist fermionic quantum anomalies related to $U(1)_\text{L}\times U(1)_\text{R}$, which cannot be realized by its bosonization $S_\text{boson}$. A typical example illustrating this ``$\mathbb{Z}_2$-killing'' effect is that $\mathbb{Z}_{2n}\times\mathbb{Z}_{2n}\subset U(1)_\text{L}\times U(1)_\text{R}$ gives a $\mathbb{Z}_{4n}$ fermionic symmetry protected trivial (SPT) phase classification, whereas $\mathbb{Z}_{2n}\times\mathbb{Z}_{2n}$ only provides a $\mathbb{Z}_{2n}$ bosonic SPT phase classification~\cite{Sule:2013aa} which is formally killed by half after bosonization. Additionally, independence on $K$ of Eq.~(\ref{charge_relation}) implies the robustness against interactions due to its topological nature and that it can be applied to Tomonaga-Luttinger liquids. 

\subsection{Bosonization: Resolution of Dirac mass condensation paradox}
It is an appropriate point to resolve the Dirac mass condensation paradox introduced before. We first restate or generalize that paradox below. 

Assume we have $N$ of $U(1)$ instantons in the spacetime $T^2$ and, for simplicity, they are localized at spacetime points $\{x_k\}_{k=1,\cdots,N}$ or
\begin{eqnarray}
\label{gauge_field}
eF_{01}(x)=\sum_{k=1}^N2\pi\delta^2(x-x_k). 
\end{eqnarray}
Let us evaluate the path-integral (P-T) expectation value of a series of Dirac mass bilinear term: 
\begin{eqnarray}
\left\langle\prod_{j=1}^M\bar{\Psi}(y_j)\Psi(y_j)\right\rangle_\text{P-T}\equiv\left\langle\prod_{j=1}^M\bar{\Psi}(y_j)\Psi(y_j)\right\rangle_\text{P-T: $S_\text{Dirac}$}
\end{eqnarray}
in which $\bar{\Psi}$ denotes the Dirac adjoint of $\Psi$ and it should be distinguished from the anti-holomorphism notation used before for $\bar{\psi}(\bar{z})$. To evaluate the above expectation value, we expand $\Psi$ and $\bar{\Psi}$ into their eigen-function of Dirac operator $D\equiv\gamma^\mu(\partial_\mu-ieA_\mu)$: $D\Psi_n=\lambda_n\Psi_n$ with
\begin{eqnarray}
\Psi=\sum_n a_n\Psi_n,\,\bar{\Psi}=\sum_n \bar{a}_n\bar{\Psi}_n,\,\,\int\bar{\Psi}_m\Psi_n=\delta_{m,n}, 
\end{eqnarray}
where $\{\bar{a}_n\}$ and $\{a_n\}$ are independent Grassmanian numbers. Then
\begin{eqnarray}
\label{bilinear}
&&\left\langle\prod_{j=1}^M\bar{\Psi}(y_j)\Psi(y_j)\right\rangle_\text{P-T}\!\!\!
\left\{\!\!\begin{array}{ll}\neq 0, &\text{ if }M\geq N\&M=N\!\!\!\!\mod2;\\=0,&\text{ otherwise}, \end{array}\right.\nonumber\\
\end{eqnarray}
where we have made use of the Atiyah-Singer index theorem which implies that the number of zero mode of Dirac operator is the instanton number $N$, and the ``mod $2$'' results from the fact that, for any $\lambda_n$ with $\Psi_n$ in the spectrum of Dirac operator $D$, we have
\begin{eqnarray}
D\gamma^3\Psi_n=-\lambda_n\gamma^3\Psi_n, 
\end{eqnarray}
in which $\{D,\gamma^3\}=0$ is made of use and $\gamma^3$ is canonically well-defined on any spin manifold. 

Then the paradox follows: if we assume the bosonization of the fermion model is $S'[A]$ defined in Eq.~(\ref{bosonization_wr}), applying the operator correspondence $\bar{\Psi}\Psi\propto\cos\varphi$, we obtain
$\left\langle\prod_{j=1}^M\bar{\Psi}(y_j)\Psi(y_j)\right\rangle_\text{P-T}\propto\left\langle\prod_{j=1}^M\cos\varphi(y_j)\right\rangle_{\text{P-T: }S'}\neq0$ generically for any $M$ value independing on $N$, 
which fails to match the fermionic statement in Eq.~(\ref{bilinear}). 
Thus the conventional gauged bosonic model $S'[A]$ or its generalization gauged WZW model is problematic and inconsistent with its presumed fermionic partner. 

We will solve the inconsistency or paradox above by our proposed bosonization $S_\text{boson}[A]$ defined in Eq.~(\ref{boson_2}). Due to the localized gauge-field configuration in Eq.~(\ref{gauge_field}), we have $S_\text{boson}[A]=S_0^\text{(b)}-i\sum_{k=1}^N\varphi(x_k)$, and thus
\begin{eqnarray}
&&\left\langle\prod_{j=1}^M\bar{\Psi}(y_j)\Psi(y_j)\right\rangle_\text{P-T}\propto\left\langle\prod_{j=1}^M\cos\varphi(y_j)\right\rangle_\text{P-T: $S_\text{boson}$}\nonumber\\
&=&\left\langle\prod_{j,k}\frac{\exp[i\varphi(y_j)]+\exp[-i\varphi(y_j)]}{2}\exp[i\varphi(x_k)]\right\rangle_\text{P-T: $S_0^\text{(b)}$}\nonumber\\
&&\left\{\begin{array}{ll}\neq 0, &\text{ if }M\geq N\&M=N\mod2;\\=0,&\text{ otherwise}, \end{array}\right.
\end{eqnarray}
which is exactly the fermionic result in Eq.~(\ref{bilinear}), and we have applied the neutrality condition for $\varphi$'s path integral upon action $S_0^\text{(b)}$. We can see that the constraint given by Atiyah index theorem on the fermionic side precisely corresponds to that by neutrality condition. Therefore, with our new bosonization $S_\text{boson}[A]$, the paradox brought by wrong $S'[A]$ has been resolved successfully. The similar argument may be also straightforward to be applied for higher symmetries with nontrivial fundamental homotopy group, e.g. $SU(N)/\mathbb{Z}_N$. 

\label{spin chain}
\section{Quantum $XXZ$ chain with twisted boundary condition}

In this part, we further apply our bosonization in background gauge field to quantum ferromagnetic $XXZ$ spin chain with an antiferromagnetic anisotropy along $z$ axis: 
\begin{eqnarray}
H_{XXZ}=-\sum_{i=1}^L\left(\sigma_i^x\sigma_{i+1}^x+\sigma_i^y\sigma_{i+1}^y+\Delta\sigma_i^z\sigma_{i+1}^z\right), 
\end{eqnarray}
with $\Delta\equiv-\cos\gamma$ and $\gamma\in[0,\pi]$ and the following twisted boundary condition (TBC)
\begin{eqnarray}
\label{tbc}
\sigma_{L+1}^x\pm i\sigma_{L+1}^y=\exp(\pm i\phi_\text{tw})(\sigma_1^x\pm i\sigma_1^y). 
\end{eqnarray}
{The effect of twisted boundary condition of XXZ Heisenberg model
has been considered in the framework of the integrable model.
Numerical calculations of this model have been achieved by the
seminal work by~\cite{ALCARAZ1988280} and the excitation spectrum has also been considered~ \cite{Nassar:1998aa}. 
The following application of our results on $XXZ$ models is also largely motivated by such numerical results, which is a consistency check. }

\begin{figure}
\begin{center}
\includegraphics[width=8cm,pagebox=cropbox,clip]{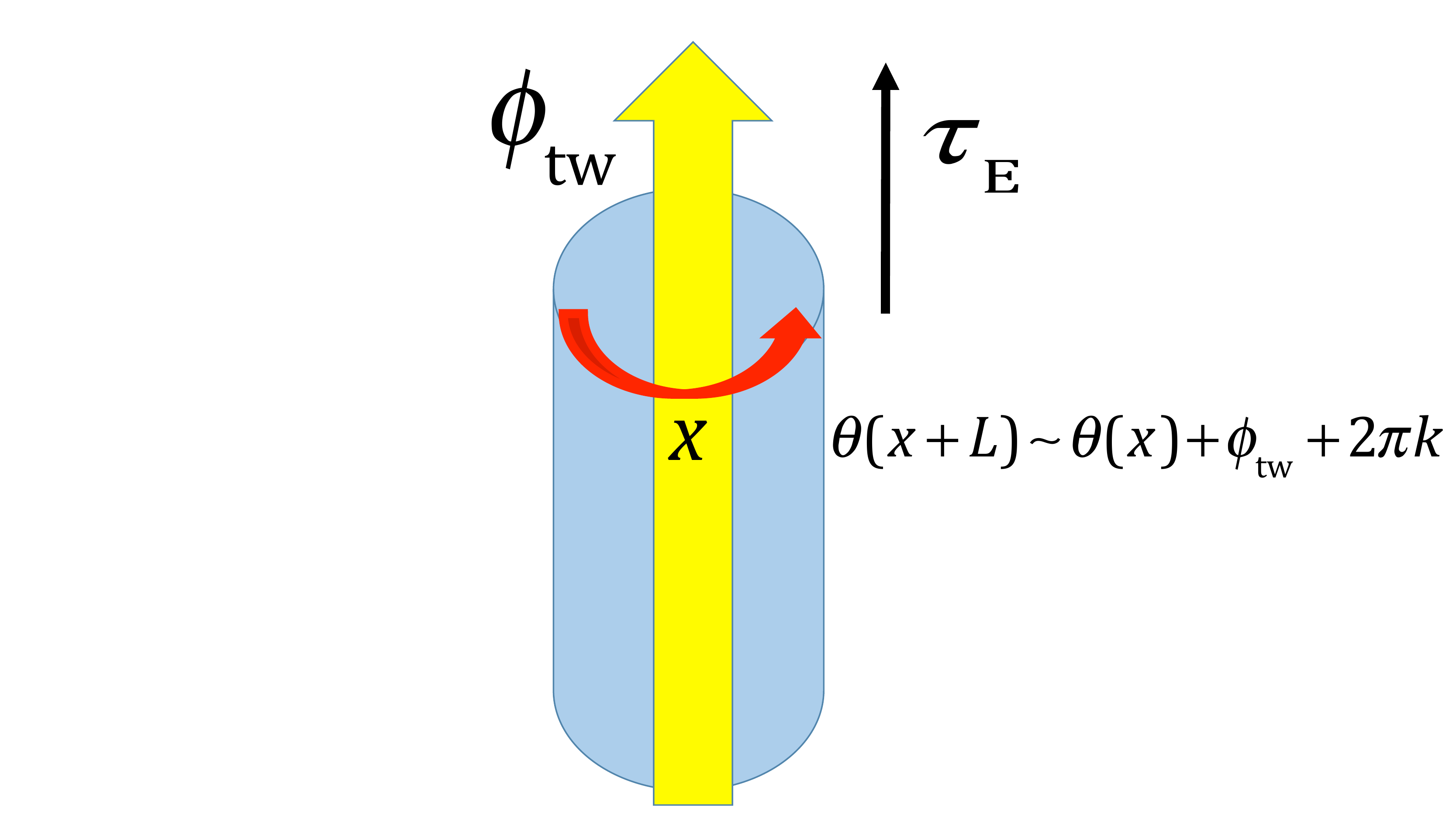}
\caption{Twisted boundary condition of free boson with a winding number $k$ in Eq.~(\ref{xxzpartition}). }
\label{spin_flux}
\end{center}
\end{figure}

In the case of periodic boundary condition (PBC) $\phi_\text{tw}=0$, the lattice model can be mapped to a spinless lattice fermionic model with interaction by Jordan-Wigner transformation whose low-energy physics is captured by the Tomonaga-Luttinger liquid~\cite{Affleck:1988aa}: 
\begin{eqnarray}
S^{0}_\text{T-L}&=&\frac{1}{8\pi K}\int d\tau_\text{E}dx\left[(\partial_{\tau_\text{E}}\varphi)^2+(\partial_x\varphi)^2\right],
\end{eqnarray}
where $\tau_\text{E}\equiv ivt$ with $K\equiv\pi/[2(\pi-\gamma)]$ and $v$ normalizing factors for Luttinger parameter $1/4\pi$ and Fermi velocity $v_\text{F}$, respectively, due to the interaction of the corresponding fermionic model. In our convention, the bosonic field $\varphi$ is always normalized so that its radius is unity: $\varphi\sim\varphi+2\pi$. The $XY$ point where $\gamma=\pi/2$ corresponds to $K=1$, exactly the free fermionic point. 

Let us assume that the minimal coupling between $\varphi$ and $A$ is renormalized by $\Gamma_K$ with $K\neq1$: 
\begin{eqnarray}
&&S_\text{T-L}[A]-S^0_\text{T-L}\nonumber\\
&=&-i\Gamma_K\frac{e}{2\pi}\left(\int_{T^2}\varphi dA-i\epsilon^{\mu\nu}\int_{\text{cycle}_\mu}\!\!d\varphi\int_{\text{cycle}_\nu}\!\!\!A\right). 
\end{eqnarray}
Since the bosonic field is angle-like with periodicity $2\pi$ and the general gauge field always satisfies $\int edA/2\pi\in\mathbb{Z}$ on torus, we need the action invariant under the redundancy of field description $\varphi\rightarrow\varphi+2\pi$ for any background field. Such an invariance requires
\begin{eqnarray}
\label{non-re}
\Gamma_K=1, 
\end{eqnarray}
since the Chern class $\int edA/2\pi=\pm1$ can be realized on torus which reflects the topological nature of this non-renormalization. 
\subsection{Low-energy partition function}
\label{TL_liquid_0}
Since the TBC in the spin-chain language, following the Jordan-Wigner transformation, is translated to the twisted boundary condition for fermion by a charge $U(1)$ phase transformation: $c_{j+L}=c_j\exp(i\phi_\text{tw})$. It can be realized by a flat background gauge field with $dA=0$ and $\beta=0$: $\phi_\text{tw}=2\pi\alpha=e\int_\text{cycle$_1$}A(x^0,x^1)$. 

Therefore, we obtain the bosonization of our twisted $XXZ$ chain as: 
\begin{eqnarray}
S_\text{T-L}[A]=S^0_\text{T-L}-i2\pi\frac{\phi_\text{tw}}{2\pi}\int_\text{cycle$_0$}\frac{\partial_{\tau_\text{E}}\varphi}{2\pi}.
\end{eqnarray}
This action is consistent with that of Kitazawa by taking the limit $q\rightarrow 0$.
It is a variation of Dotsenko-Fateev Coulomb gas model \cite{Dotsenko:1985aa}.
Hence we can understand the flux insertion of Dirac fermion as the background charge insertion of free boson.  
We see that its partition function is exactly Eq.~(\ref{flatpartition}) with $\alpha=\phi_\text{tw}/2\pi$ and $\beta=0$: 
\begin{eqnarray}
\label{xxzpartition}
Z_\text{TL}(K,\phi_\text{tw})\!\!&=&\!\!\frac{1}{|\eta(\tau)|^2}\!\!\!\sum_{n,k\in\mathbb{Z}}\!q^{\frac{K}{2}\left(k+\frac{n}{2K}+\frac{\phi_\text{tw}}{2\pi}\right)^2}\bar{q}^{\frac{K}{2}\left(k-\frac{n}{2K}+\frac{\phi_\text{tw}}{2\pi}\right)^2}, \nonumber\\
\end{eqnarray}
which can be seen as the $\theta$-field twisting as FIG.~(\ref{spin_flux}).

\subsection{Short review of XXZ spin chain with twisted boundary condition}

The field theoretic analysis and the
relation of XXZ spin chains to quantum group were considered~\cite{Faddeev:1995aa,Pallua:1996aa}.
However, except for the work by Kitazawa~\cite{Fukui:1996aa, Kitazawa:1997aa}, no one has
ever considered this effect as that of background gauge field.
The reason why almost no one considers such problem may be related to
the interpretations of background gauge fields.
The interpretation of such gauge transformation for condensed matter physicists
is different from that of high energy physicists in the sense that the former
does not gauge out the background gauge field.
Kitazawa has numerically and phenomenologically shown
this effect can be described by the effect of the background
charge of free boson.
At this stage, it is difficult to understand the equivalence of fermion with flux and boson with
background charge. Hence the more systematic derivation of his results is desired.

The more combinatorial approach on this phenomena was considered in the context of
polynomial and integrable field theory. Combinatorial equivalence of this boson-fermion correspondence is
called Rojers-Ramanujan identity \cite{Welsh:2006aa, Welsh:2005aa}.
This identity relates the character of minimal model to $q$ deformed fermionic sum.

\subsection{Correspondences between quantum $XXZ$ chain and Potts model}
In this part, we will see how the partition function of the low-energy twisted $XXZ$ chain Eq.~(\ref{xxzpartition}) can give properties of the thermal operator in $Q$-state Potts model for $Q\leq4$. 
Correspondence between twisted free boson and  Potts model has been considered by
\cite{Di-Francesco:1987aa}. In other words, our formalism is a spin chain version of this work.

The quantum $Q$ state Potts model takes the form as
\begin{eqnarray}
H^{\text{P}}_\text{Q-Potts}=-\sum_{i=1}^{L}\sum_k^{Q-1}\Omega_i^k-\sum_{i=1}^L\sum_{k=1}^{Q-1}R_i^kR_{i+1}^{Q-k}, 
\end{eqnarray}
with PBC: $R_{L+1}=R_1$ denoted by ``P'' in the superscript in ``$H^{\text{P}}_\text{Potts}$'' and the $Z(q)$ algebra $(\omega\equiv\exp(i2\pi/Q))$ is satisfied by $R$'s and $Q$'s: 
\begin{eqnarray}
\Omega_iR_i=\omega^{-1}R_i\Omega_i;\,\,\,\Omega_iR_i^\dagger=\omega R_i^\dagger\Omega_i; \,\,\,\Omega_i^Q=R^Q_i=1. 
\end{eqnarray}
$H^{\text{P}}_\text{Q-Potts}$ can be diagonalized into blocks labelled by $H^{\text{P},q}_\text{Q-Potts}$ with $\prod_{i=1}^L\Omega_i=\omega^{q}$: 
\begin{eqnarray}
H^{\text{P}}_\text{Q-Potts}=\text{diag}\left[H^{\text{P},0}_\text{Potts},H^{\text{P},1}_\text{Q-Potts},\cdots,H^{\text{P},Q-1}_\text{Q-Potts}\right]. 
\end{eqnarray}

\subsubsection{Thermal operator: $\varepsilon$}
It has been proven that, on the operator level, the following correspondence between the ground-state sector of Potts model and twisted $XXZ$ chain holds up to an irrelevant constant shift: 
\begin{eqnarray}
\label{pottsxxz}
H^{\text{P},0}_\text{Q-Potts}=H_\text{XXZ}(\gamma,\phi_\text{tw}=2\gamma)
\end{eqnarray}
by an appropriate normalization of coupling constant $J_\text{XXZ}$ and setting $\gamma=\arccos(\sqrt{Q}/2)$~\cite{ALCARAZ1988280}. 

By finite-size scaling of correlation length, the thermal operator ``$\varepsilon$'' of $Q$-state Potts model lies exactly at the first excited state of the sub-Hamiltonian $H^\text{P,0}_\text{Q-Potts}$. By the correspondence in Eq.~(\ref{pottsxxz}) above, we see that the conformal properties of $\varepsilon$ can be extracted out by the partition in Eq.~(\ref{xxzpartition}): 
\begin{eqnarray}
\label{xxzpottspartition}
&&Z_\text{TL}\left(\frac{\pi}{2(\pi-\gamma)},2\gamma\right)\nonumber\\
&=&\frac{1}{|\eta(\tau)|^2}\sum_{n,k\in\mathbb{Z}}q^{\frac{\pi}{4(\pi-\gamma)}\left[k+\frac{n(\pi-\gamma)}{\pi}+\frac{\gamma}{\pi}\right]^2}\bar{q}^{\frac{\pi}{4(\pi-\gamma)}\left[k-\frac{n(\pi-\gamma)}{\pi}+\frac{\gamma}{\pi}\right]^2}\!\!, \nonumber\\
\end{eqnarray}
from which we can read off the conformal anomaly defined as the lowest conformal weight of the critical $Q$-state Potts model after setting $\gamma=\arccos(\sqrt{Q}/2)$: 
\begin{eqnarray}
c=1-\frac{6\arccos(\sqrt{Q}/2)^2}{\pi(\pi-\arccos(\sqrt{Q}/2))}, 
\end{eqnarray}
and the conformal weight of $\varepsilon$ by setting its first excited energy eigenstate labelled by $(k,n)=(1,0)$: 
\begin{eqnarray}
\Delta_\varepsilon&=&\bar{\Delta}_\varepsilon=\frac{\pi+2\arccos(\sqrt{Q}/2)}{4(\pi-\arccos(\sqrt{Q}/2))}, 
\end{eqnarray}
where $\Delta_\varepsilon$ and $\bar{\Delta}_\varepsilon$ are, respectively, holomorphic and anti-holomorphic conformal dimensions of $\epsilon$. These properties exactly match those of conformal field theories of low-energy $Q$-state Potts model. 

\subsubsection{Order operator $\sigma$}
For the other operators such as order parameter and para-fermion operator, their location are only empirically identified in the spectrum of twisted $XXZ$ spin chains with other twisted boundary conditions. More specifically, the order parameter $\sigma$ of Potts model can be found in the spectrum of $XXZ$ chain with twisted angle as $\phi_\text{tw}=\pi$ with the partition function as
\begin{eqnarray}
\label{xxzorderpartition}
&&Z_\text{TL}\left(\frac{\pi}{2(\pi-\gamma)},\pi\right)\nonumber\\
&=&\frac{1}{|\eta(\tau)|^2}\!\!\sum_{n,k\in\mathbb{Z}}q^{\frac{\pi}{4(\pi-\gamma)}\left[k+\frac{n(\pi-\gamma)}{\pi}+\frac{1}{2}\right]^2}\bar{q}^{\frac{\pi}{4(\pi-\gamma)}\left[k-\frac{n(\pi-\gamma)}{\pi}+\frac{1}{2}\right]^2}. \nonumber
\end{eqnarray}
The conformal dimensions of $\sigma$ is empirically determined by the lowest energy eigenstate of $n=0$ sector of $XXZ$ Hamiltonian, namely $(k,n)=(0,0)$: 
\begin{eqnarray}
\Delta_\sigma+\frac{\gamma^2}{4\pi(\pi-\gamma)}&=&\frac{\pi}{16(\pi-\gamma)}\\
\bar{\Delta}_\sigma+\frac{\gamma^2}{4\pi(\pi-\gamma)}&=&\frac{\pi}{16(\pi-\gamma)}, 
\end{eqnarray}
which are solved as $\Delta_\sigma=\bar{\Delta}_\sigma=({\pi^2-4\gamma^2})/[16\pi(\pi-\gamma)]$. 
\subsection{Parafermion operators}
Numerically, the parafermion operator with its spin as $\tilde{Q}/Q$ with $\tilde{Q}=1,2,\cdots,Q-1$ can find its location in the lowest energy eigenstate of $n=1$ sector of the spectrum of $XXZ$ chain with twisted angle as $\phi_\text{tw}=2\pi\tilde{Q}/Q$. To obtain its conformal properties, we write down the corresponding partition function of $XXZ$ chain: 
\begin{eqnarray}
\label{xxzpottspartition}
&&Z_\text{TL}\left(\frac{\pi}{2(\pi-\gamma)},\frac{2\pi\tilde{Q}}{Q}\right)\nonumber\\
&=&\frac{1}{|\eta(\tau)|^2}\!\!\sum_{n,k\in\mathbb{Z}}\!q^{\frac{\pi}{4(\pi-\gamma)}\left[k+\frac{n(\pi-\gamma)}{\pi}+\frac{\tilde{Q}}{Q}\right]^2}\!\bar{q}^{\frac{\pi}{4(\pi-\gamma)}\left[k-\frac{n(\pi-\gamma)}{\pi}+\frac{\tilde{Q}}{Q}\right]^2}, \nonumber
\end{eqnarray}
which implies, after $(k,n)=(0,1)$ is extracted out, 
\begin{eqnarray}
\Delta_\text{pf}+\frac{\gamma^2}{4\pi(\pi-\gamma)}&=&\frac{\pi-\gamma}{4\pi}+\frac{\tilde{Q}}{2Q}+\frac{\pi\tilde{Q}^2}{4Q^2(\pi-\gamma)};\\
\bar{\Delta}_\text{pf}+\frac{\gamma^2}{4\pi(\pi-\gamma)}&=&\frac{\pi-\gamma}{4\pi}-\frac{\tilde{Q}}{2Q}+\frac{\pi\tilde{Q}^2}{4Q^2(\pi-\gamma)}, 
\end{eqnarray}
which are solved as
\begin{eqnarray}
\Delta_\text{pf}&=&\frac{\pi-\gamma}{4\pi}+\frac{\pi^2\tilde{Q}^2-\gamma^2Q^2}{4\pi Q^2(\pi-\gamma)}+\frac{\tilde{Q}}{2Q},\\
\bar{\Delta}_\text{pf}&=&\frac{\pi-\gamma}{4\pi}+\frac{\pi^2\tilde{Q}^2-\gamma^2Q^2}{4\pi Q^2(\pi-\gamma)}-\frac{\tilde{Q}}{2Q}, 
\end{eqnarray}
which exactly imply the spin as $\Delta_\text{pf}-\bar{\Delta}_\text{pf}=\tilde{Q}/Q$, namely the spin of the parafermion operator. 

\section{Conclusion}
\label{conclusions}
In this paper, we proposed the new bosonization of Dirac fermion coupled with background gauge field.
Our formalism can be thought of the natural extension of existing bosonization.
It is consistent with the global chiral anomaly of Dirac fermion and with the operator correspondence of $U(1)$ charge current.
Moreover it can describe the numerical and Bethe ansatz results
of XXZ chain with twisted boundary conditions.

As we have discussed, the functional bosonization is insufficient for $(1+1)$-dimensional due to nontrivial fundamental homotopy of underlying space-time manifold.
Hence it is natural to arrive at that the perturbations which suppress the topologically nontrivial field-configuration
sectors, e.g. nonzero winding in $U(1)$ boson case, may assure the validity of functional bosonization.
Otherwise, the evaluation of the contribution of each topological sector is necessary for using the functional bosonization.

Our formalism clearly suggests the relationship between the fermionic system with
twisted boundary condition and the bosonic system with background charge~\cite{Furuya:2019aa}.
Moreover the generalization of $U(1)$ gauge symmetry to $SU(N)/\mathbb{Z}_N$ seems to be
straightforward.
Hence we conjecture multi-component fermion coupled with gauge field
can be bosonized to multiple bosons with background charge.
The later theory includes a wide variety of CFTs like Toda field theory and
Coulomb gas representation of WZW models.
In that sense, twisted XXZ chain and its field theoretic analog shows 
the correspondence between minimal CFT and twisted fermion.
For more general application, we can understand the result
by A Klumper \cite{Klumper:1991aa} in our formalism if we think of spin $S$ chain
with twisted boundary condition as parafermion and free boson 
with background charge \cite{Konno:1993aa}. For the simplest example, we can identify
spin 1 chain as one Dirac fermion and Majorana fermion \cite{Di-Francesco:1988aa}.
If we add the twist to this Dirac fermion and consider the bosonization, 
we obtain free boson with background charge and Majorana fermion.
Hence this model should reproduce central charge of $N=1$ minimal conformal
field theory.

Related to this observation, the relation between the insensitivity of the twist angle of the gapped system and
massive integrable model derived from CFT is quite mysterious because 
CFT is not insensitive to twistings. The integrable perturbations and their flow were
researched by using the thermodynamic Bethe ansatz, hence further research on
this insensitivity of twist may result in some universal RG flows under twist.
The TQFT/CFT correspondence may shed new light for this problem
because Chern-Simons-BF theory is known to become TQFT.

Moreover, as one of the authors has shown, there exist some difficulties to
express nonlocal lattice object related to large gauge transformation \cite{PhysRevB.97.165133}.
Such an object is the gapless version of the indicator and it should be described by low energy field
theory. Our formalism is supposed to explain such phenomena more consistently.

\section{Acknowledgements}
The authors thank Ryan Thorngren for discussion and reminder on symmetry extension brought by bosonization. 
Yuan Yao is grateful to David Tong for useful discussions on bosonizations and Linhao Li for a discussion on the CS-WZW duality. 
Yoshiki Fukusumi (YF) thanks comments and discussion of Chihiro Matsui, Haruki Watanabe, and Raoul Santachiara.
YF also thanks Ryohei Kobayashi, Yuya Nakagawa, and Masaki Oshikawa for sharing the related problems.

\appendix*
\begin{widetext}
\section{Derivations and further applications}
\subsection{Derivations of $Z_\text{boson}[K,A_\text{flat}]$ in Eq.~(\ref{flatpartition})}
\label{z_flat}
Let us calculate the partition function associated with $S_\text{boson}[K;\alpha,\beta]$: 
\begin{eqnarray}
Z_\text{boson}[A_\text{flat}]=Z_\text{winding-free}\cdot Z_\text{w}, 
\end{eqnarray}
in which
\begin{eqnarray}
Z_\text{winding-free}=\frac{1}{\sqrt{2\tau_2 K}|\eta(\tau)|^2}, 
\end{eqnarray}
and the winding-number contribution is
\begin{eqnarray}
\label{winding}
Z_\text{w}&=&\sum_{n,n'}\exp\left\{-\frac{\pi}{2K}\left[\frac{1}{\tau_2}(n'-\tau_1n)^2+\tau_2n^2\right]\right\}\exp\left[-i2\pi(n\beta-n'\alpha)\right]\nonumber\\
&=&\sum_{n,n'}\exp\left\{-\pi\left[\frac{1}{2\tau_2K}(n')^2\right]+\left[\pi\frac{\tau_1}{\tau_2K}n+i2\pi\alpha\right]n'\right\}\exp\left[-\frac{\pi}{2}\frac{|\tau|^2}{\tau_2K}n^2-i2\pi n\beta\right]. 
\end{eqnarray}
We could use the following Poisson resummation formula
\begin{eqnarray}
\sum_{n'\in\mathbb{Z}}\exp\left[-\pi a{n'}^2+bn'\right]=\frac{1}{\sqrt{a}}\sum_{k\in\mathbb{Z}}\exp\left[-\frac{\pi}{a}\left(k+\frac{b}{2\pi i}\right)^2\right], 
\end{eqnarray}
and obtain 
\begin{eqnarray}
\label{zwpartition}
Z_\text{w}&=&\sqrt{2\tau_2K}\sum_{n,k}\exp\left[-\frac{\pi}{2}\frac{|\tau|^2}{\tau_2K}n^2-i2\pi n\beta\right]\exp\left\{-2\pi\tau_2K\left[k+\frac{1}{2\pi i}\left(\pi\frac{\tau_1}{\tau_2K}n+i2\pi\alpha\right)\right]^2\right\}\nonumber\\
&=&\sqrt{2\tau_2K}\sum_{n,k}\exp(-i2\pi n\beta)q^{\frac{K}{2}\left(k+\frac{n}{2K}\right)^2+\alpha K\left(k+\frac{n}{2K}\right)}\bar{q}^{\frac{K}{2}\left(k-\frac{n}{2K}\right)^2+\alpha K\left(k-\frac{n}{2K}\right)}(q\bar{q})^\frac{\alpha^2K}{2}, 
\end{eqnarray}
where $q\equiv\exp(i2\pi\tau)$ and $\bar{q}\equiv\exp(-i2\pi\bar{\tau})$. 

Combining the forms of $Z_\text{winding-free}$ with $Z_\text{w}$, we obtain
\begin{eqnarray}
\label{flatpartition_1}
&&Z_\text{boson}[K,A_\text{flat}]=Z_\text{winding-free}\cdot Z_\text{w}\\
&=&\sum_{n,k}\exp(-i2\pi n\beta)q^{\frac{K}{2}\left(k+\frac{n}{2K}+\alpha\right)^2}\bar{q}^{\frac{K}{2}\left(k-\frac{n}{2K}+\alpha\right)^2}/|\eta(\tau)|^2. \nonumber
\end{eqnarray}

\subsection{Applications}
In this part, we will apply our bosonization to several real systems. The space-time will be taken to be $(t,x)$ rather than the Euclidean one. 
The readers should keep in mind the results of this section is derived only from bosonization but not from exact methods.
The results of this section may be verified by numerical simulation or Bethe ansatz of the spin chain~\cite{De-Luca:2014aa,Nakagawa:2016aa}.

\subsubsection{Spinless electronic system in the background electromagnetic field}

Let us consider the following spinless fermion on one-dimensional chain: 
\begin{eqnarray}
\mathscr{H}=-\frac{J_{XY}}{2}\left(\sum_{j=1}^{L/a}c_{j+1}^\dagger\exp[iA_1(j)a]c_j+\text{h.c.}+A_0(j)c_j^\dagger c_j\right), 
\end{eqnarray}
where $a$ is the lattice constant and $J_{XY}$ is the hopping amplitude, which can also be the antiferromagnetic exchange coupling constant in the $XY$ model having the same fermionic representation by $c_j$'s and $c_j^\dagger$'s. We could go to the low-energy limit and it simply gives the Dirac fermion coupled with the background gauge field. Then we bosonize the theory with the following operator correspondences between lattice operators and bosonic ones: 
\begin{eqnarray}
\label{density}
\delta\rho\leftrightarrow-\frac{1}{2\pi}\partial_x\varphi&=&\frac{i}{2\pi}(\partial_w-\partial_{\bar{w}})\varphi(w,\bar{w})\nonumber\\
J\leftrightarrow+\frac{1}{2\pi}\partial_t\varphi&=&i\frac{v_\text{F}}{2\pi}(\partial_w+\partial_{\bar{w}})\varphi(w,\bar{w}), 
\end{eqnarray}
where $w=\tau_\text{E}-ix$, $\bar{w}=\tau_\text{E}+ix$, $\tau_\text{E}\equiv iv_\text{F}t$ and $\delta\rho\equiv(\rho-\nu)$ with filling factor $\nu$ and $v_\text{F}$ the Fermi velocity. It should be noted that the Luttinger parameter is still $1/4\pi$ since the coefficient $v_\text{F}$ of Minkowskian Langrangian density will be absorbed into the integration measure $\int d\tau_\text{E}dx$ of action. 

For later use, we do a conformal mapping: $z=\exp(2\pi w/L)$ where $L$ is the circumference of the cylinder. Then, 
\begin{eqnarray}
J&=&i\frac{v_\text{F}}{2\pi}\left[(\partial_wz)\partial_z+(\partial_{\bar{w}}\bar{z})\partial_{\bar{z}}\right]\varphi(z,\bar{z})\nonumber\\
&=&i\frac{v_\text{F}}{L}\left(z\partial_z+\bar{z}\partial_{\bar{z}}\right)\varphi(z,\bar{z}), 
\end{eqnarray}
and, similarly, 
\begin{eqnarray}
\delta\rho=\frac{i}{L}\left(z\partial_z-\bar{z}\partial_{\bar{z}}\right)\varphi(z,\bar{z}). 
\end{eqnarray}
We have omitted the oscillating parts of $J$ and $\delta\rho$ since they do not contribute to the correlation function we are interested in the rest sections. 

\subsubsection{Time-ordered correlation function}
Let us denote the time evolution operator defined by our system including (time-dependent) interactions by $U(+\infty,-\infty)$. The quantum average of an observable $\hat{X}$ at time $t$ is
\begin{eqnarray}
\langle X(t)\rangle_\text{QM}\equiv\langle0|U^\dagger(t,-\infty)\hat{X}U(t,-\infty)|0\rangle. 
\end{eqnarray}
To relate it with the path integral language, 
we need to find an action $\tilde{S}$ and its time revolution operator $\tilde{U}$, up to a phase factor $\exp(il)$, satisfying, 
\begin{eqnarray}
\label{condition_0}
&&\exp(il)|0\rangle=\tilde{U}(+\infty,-\infty)|0\rangle,\nonumber\\
&&\left.\tilde{U}(t',-\infty)\right|_{t'\leq t}=U(t',-\infty). 
\end{eqnarray}
Therefore, 
\begin{eqnarray}
\label{qmave}
\langle X(t)\rangle_\text{QM}&=&\langle0|\tilde{U}^\dagger(t,-\infty)\hat{X}\tilde{U}(t,-\infty)|0\rangle\nonumber\\
&=&\langle0|\tilde{U}(-\infty,+\infty)\tilde{U}(+\infty,t)\hat{X}\tilde{U}(t,-\infty)|0\rangle\nonumber\\
&=&\exp(-il)\langle0|\tilde{U}(+\infty,t)\hat{X}\tilde{U}(t,-\infty)|0\rangle\nonumber\\
&=&\frac{\langle0|\tilde{U}(+\infty,t)\hat{X}\tilde{U}(t,-\infty)|0\rangle}{\langle0|\tilde{U}(+\infty,-\infty)|0\rangle}, 
\end{eqnarray}
which is exactly a path integral representation after a Wick rotation $t\rightarrow t\exp(-i\epsilon)$ and $\pm\infty\rightarrow\pm\infty\exp(-i\epsilon)$ with $\epsilon\rightarrow0^+$, or equivalently, one could rotate it to Euclidean space-time that we will do next. 

\subsubsection{Spatially uniform electromagnetic pulse: $eF_{01}=2\pi\delta(t-t_0)/L$}
Let us consider a case that the background gauge field is uniform in the spatial component while a pulse in the temporal direction: 
\begin{eqnarray}
eF_{01}=2\pi\delta(t-t_0)/L, 
\end{eqnarray}
where $F_{01}$ has a proper quantization that $\int F=\int dtdx\,F_{01}\in2\pi\mathbb{Z}/e$ and it has been transformed back to a Minkowskian tensor, while $\int F$ is coordinate-independent. 

Let us first map the system on the cylinder onto the complex plane by $z=\exp(w)$, and consider the expectation value of $\partial_w\varphi(w,\bar{w})$ and $\partial_{\bar{w}}\varphi(w,\bar{w})$. 
Then, if $t<t_0$, the observable $\delta\rho(t)$ and $J(t)$ cannot be influenced by the future pulse, or equivalently speaking, we simply take $\tilde{S}=S_0^\text{(b)}$ with the free time evolution $\tilde{U}=\tilde{U}_0$ which obviously satisfies Eq.~(\ref{condition_1}). 

Next, when $t>t_0$, we should fix the $\tilde{U}$ and its action $\tilde{S}$ satisfying Eq.~(\ref{condition_0}). Let us take the following action: 
\begin{eqnarray}
\tilde{S}(t_1)&=&S_0-\int_{t_1>t}\frac{dx_1}{L}\varphi(t_1,x_1)+\int_{t_2=t_0}\frac{dx_2}{L}\varphi(t_2,x_2). 
\end{eqnarray}
Obviously, $\langle0|\tilde{U}(+\infty,-\infty)|0\rangle$ is converging to $1$ as $t_1$ is approaching $t=t_0$ from left (although, by definition $t_1>t>t_0$, we are still free to extend its domain). Let us show that it satisfies Eq.~(\ref{condition_0}). Indeed, 
\begin{eqnarray}
\label{condition_1}
&&\frac{d}{dt_1}\langle0|\tilde{U}(+\infty,-\infty)|0\rangle\nonumber\\
&=&\int\mathscr{D}\varphi\left\{\exp\left[-i\int_{t_1>t}\frac{dx_1}{L}\varphi(t_1,x_1)+i\int_{t_2=t_0} \frac{dx_2}{L}\varphi(t_2,x_2)+iS_0\right]\left[\int_{t_1>t}\frac{dx_1}{iL}\partial_{t}\varphi(t_1,x_1)\right]\right\}\nonumber\\
&=&\frac{\left\langle0\left|\tilde{U}(+\infty,t_1)\left[\int_{t_1>t}\frac{dx_1}{iL}\partial_{t}\varphi(t_1,x_1)\right]\tilde{U}(t_1,-\infty)\right|0\right\rangle}{\langle0|\tilde{U}(+\infty,-\infty)|0\rangle}\cdot\langle0|\tilde{U}(+\infty,-\infty)|0\rangle\nonumber\\
&=&-iv_\text{F}\frac{2\pi}{L}\langle0|\tilde{U}(+\infty,-\infty)|0\rangle, 
\end{eqnarray}
which solves as
\begin{eqnarray}
\label{phase}
\langle0|\tilde{U}(+\infty,-\infty)|0\rangle=\exp\left[-iv_\text{F}\frac{2\pi}{L}(t_1-t_0)\right]. 
\end{eqnarray}
namely $\exp(il)=\exp\left[-i{2\pi}v_\text{F}(t_1-t_0)/L\right]$, which is scaling dependent as seen below. It implies that Eq.~(\ref{condition_0}) is satisfied and Eq.~(\ref{condition_1}) will be proven within Sec.~\ref{derivations}. A physical interpretation of Eq.~(\ref{condition_1}) is straightforward that the wave function does not change against the rapid pulse. What is more is that the wave function is indeed the first excited energy eigenstate of the Hamiltonian after the pulse. The excited energy, relative to the ground state, is
\begin{eqnarray}
\Delta E&=&v_\text{F}\frac{2\pi}{L}, 
\end{eqnarray}
which exactly produces the dynamical phase matching Eq.~(\ref{phase}): 
\begin{eqnarray}
\exp\left[-i\Delta E(t_1-t_0)\right]=\exp\left[-iv_\text{F}\frac{2\pi}{L}(t_1-t_0)\right]=\exp(il). 
\end{eqnarray}
However, by accident, the excited states with energy $\Delta E$ are degenerate since one single oscillating mode and a unit electromagnetic $U(1)$ phase winding mode are both of that energy level. We will see, by use of duality mappings in Sec. \ref{momentum_pulse_sec}, the excited state after the pulse should be winding mode rather than an oscillator. 

We also calculate the following quantity, 
\begin{eqnarray}
\label{qmw}
&&\left.\langle\partial_w\varphi(w,\bar{w})\rangle_\text{QM}\right|_{w=it-ix,\bar{w}=it+ix}\nonumber\\
&=&\frac{\langle0|\tilde{U}(+\infty,t)\partial_w\varphi(-ix,+ix)\tilde{U}(t,-\infty)|0\rangle}{\langle0|\tilde{U}(+\infty,-\infty)|0\rangle}\nonumber\\
&=&\frac{\int\mathscr{D}\varphi\left\{\exp\left[-i\int_{t_1>t}\frac{dx_1}{L}\varphi(t_1,x_1)+i\int_{t_2=t_0} \frac{dx_2}{L}\varphi(t_2,x_2)+iS_0\right]\left[\partial_w\varphi(w,\bar{w})\right]\right\}}{\int\mathscr{D}\varphi\exp\left[-i\int_{t_1>t}\frac{dx_1}{L}\varphi(t_1,x_1)+i\int_{t_2=t_0} \frac{dx_2}{L}\varphi(t_2,x_2)+iS_0\right]}\nonumber\\
&=&\frac{\int\mathscr{D}\varphi\left\{\exp\left[-i\int_{|z_1|>|z|} \frac{d\theta_1}{2\pi}\varphi(z_1,\bar{z}_1)+i\int_{|z_2|=\exp(\tau_{\text{E}0})} \frac{d\theta_2}{2\pi}\varphi(z_2,\bar{z}_2)-S_0\right]\left[\frac{2\pi}{L}z\partial_z\varphi(z,\bar{z})\right]\right\}}{\int\mathscr{D}\varphi\exp\left[-i\int_{|z_1|>|z|} \frac{d\theta_1}{2\pi}\varphi(z_1,\bar{z}_1)+i\int_{|z_2|=\exp(\tau_{\text{E}0})} \frac{d\theta_2}{2\pi}\varphi(z_2,\bar{z}_2)-S_0\right]}\nonumber\\
&=&-i, 
\end{eqnarray}
where we have done the conformal transformation from $(w,\bar{w})$ coordinate to $(z,\bar{z})$ by $z=\exp(2\pi w/L)$ and $\bar{z}=\exp(2\pi\bar{w}/L)$ inducing $\partial_w\varphi\rightarrow(\partial_wz)\!\cdot\!\partial_z\varphi(z,\bar{z})=2\pi z\partial_z\varphi(z,\bar{z})/L$, and we have define $\theta_{1,2}\equiv\arg(z_{1,2})$ and $\tau_{\text{E}0}=it_0$ treated as a real number in the radial quantization. On the second line of Eq.~(\ref{qmw}), we write $\partial_w\varphi(w,\bar{w})$ in its Schr\"{o}dinger representation which is the reason we have set its time variable to be zero: $\partial_w\varphi(-ix,+ix)$. 

Similarly, for the anti-holomorphic component, 
\begin{eqnarray}
\label{qmwbar}
&&\left.\langle\partial_{\bar{w}}\varphi(w,\bar{w})\rangle_\text{QM}\right|_{w=it-ix,\bar{w}=it+ix}\nonumber\\
&=&\frac{\langle0|\tilde{U}(+\infty,t)\partial_{\bar{w}}\varphi(-ix,+ix)\tilde{U}(t,-\infty)|0\rangle}{\langle0|\tilde{U}(+\infty,-\infty)|0\rangle}\nonumber\\
&=&\frac{\int\mathscr{D}\varphi\left\{\exp\left[-i\int_{|z_1|>|z|} \frac{d\theta_1}{2\pi}\varphi(z_1,\bar{z}_1)+i\int_{|z_2|=\exp(\tau_{\text{E}0})} \frac{d\theta_2}{2\pi}\varphi(z_2,\bar{z}_2)-S_0\right]\left[\frac{2\pi}{L}\bar{z}\partial_{\bar{z}}\varphi(z,\bar{z})\right]\right\}}{\int\mathscr{D}\varphi\exp\left[-i\int_{|z_1|>|z|} \frac{d\theta_1}{2\pi}\varphi(z_1,\bar{z}_1)+i\int_{|z_2|=\exp(\tau_{\text{E}0})} \frac{d\theta_2}{2\pi}\varphi(z_2,\bar{z}_2)-S_0\right]}\nonumber\\
&=&-i\frac{2\pi}{L}, 
\end{eqnarray}
where, again we set the time variable of $\partial_{\bar{w}}\varphi(w,\bar{w})$ on the second line of Eq.~(\ref{qmwbar}) to be zero since it is written in its Schr\"{o}dinger representation, and also $\bar{z}=\exp(2\pi\bar{w}/L)$ induces $\partial_{\bar{w}}\varphi\rightarrow(\partial_{\bar{w}}\bar{z})\!\!\cdot\!\!\partial_{\bar{z}}\varphi(z,\bar{z})=2\pi\bar{z}\partial_{\bar{z}}\varphi(z,\bar{z})/L$. 

We will derive Eqs.~(\ref{qmw},\ref{qmwbar}) later in Sec.~\ref{derivations}. 

Combining the calculations above with Eq.~(\ref{density}), we arrive at that
\begin{eqnarray}
\langle\delta\rho(t,x)\rangle&=&0,\nonumber\\
\langle J(t,x)\rangle&=&\frac{2\Theta(t-t_0)}{L}v_\text{F}, 
\end{eqnarray}
where $\Theta(x)$ is the step function only nonzero and valued $1$ when its argument is positive. 
This result is also exactly expected by charge-pumping argument that
\begin{eqnarray}
\label{momentum_change}
\langle P_\text{lattice}\rangle=\Theta(t-t_0)\frac{2\pi\nu}{a}\mod2\pi/a, 
\end{eqnarray}
after noticing the relation (c.f. Eq.~(\ref{momentum_pump_0}))
\begin{eqnarray}
P_\text{lattice}=\int dxk_\text{F}Jv^{-1}_\text{F}\mod2\pi/a, 
\end{eqnarray}
with $k_\text{F}=\pi\nu/a$ the Fermi momentum. 
\subsubsection{Comparison with $S'[A]$: wrong results by $S'[A]$}
If we used the previous (wrong) bosonization action $S'[A]=S^{(b)}_\text{0}+i\int\frac{e}{2\pi}A_\nu\epsilon^{\mu\nu}\partial_{\mu}\varphi$, we would have arrived at the wrong result that $J$ still had a zero expectation value after the external pulse, e.g. with the following choice of gauge as $\epsilon\rightarrow0^+$: 
\begin{eqnarray}
\left\{\begin{array}{ll}e\tilde{A}^\text{I}_0=0,\,e\tilde{A}^\text{I}_1=\frac{2\pi}{\epsilon}(t-t_0),&\text{if }(t,x)\in U_\text{I};\\
e\tilde{A}^\text{II}_0=e\tilde{A}^\text{II}_1=0, &\text{if }(t,x)\in U_\text{II}. \end{array}\right.
\end{eqnarray}
Here $U_\text{I}\equiv(t_0-\epsilon/2,t_0+\epsilon/2)\times S^1$ and $U_\text{II}=\bar{U}_\text{I}$ is its complement with $S^1$ the spatial component. It gives an obviously incorrect action that $S'[\tilde{A}]=S_0$, which produces vanishing $\langle J\rangle_\text{QM}$ inconsistent with the charge-pumping argument. 

Thus, from this example, one can see that our bosonization is necessary to produce the physically making-sense results when we have a topologically non-trivial external electromagnetic field. 

\subsection{Derivations of Eqs.~(\ref{qmw},\ref{qmwbar}) and Eq.~(\ref{condition_1})}
\label{derivations}
In this part, we will perform a detailed calculation on Eqs.~(\ref{qmw},\ref{qmwbar}) and Eq.~(\ref{condition_1}). 

We define $C_1\coloneqq\{|z_1>|z|\}$ and $C_2\coloneqq\{|z_2|=\exp(\tau_{\text{E}0})\}$ and first
\begin{eqnarray}
&&\frac{\int\mathscr{D}\varphi\left\{\exp\left[-i\int_{C_1} \frac{d\theta_1}{2\pi}\varphi(z_1,\bar{z}_1)+i\int_{C_2} \frac{d\theta_2}{2\pi}\varphi(z_2,\bar{z}_2)-S_0\right]\left[\frac{2\pi}{L}z\partial_z\varphi(z,\bar{z})\right]\right\}}{\int\mathscr{D}\varphi\exp\left[-i\int_{C_1} \frac{d\theta_1}{2\pi}\varphi(z_1,\bar{z}_1)+i\int_{C_2} \frac{d\theta_2}{2\pi}\varphi(z_2,\bar{z}_2)-S_0\right]}\nonumber\\
&=&\frac{\int\mathscr{D}\varphi\left\{\exp\left[-i\int_{C_1} \frac{d\theta_1}{2\pi}\varphi(z_1,\bar{z}_1)+i\int_{C_2} \frac{d\theta_2}{2\pi}\varphi(z_2,\bar{z}_2)-S_0\right]\left[\frac{2\pi}{L}z\partial_z\varphi(z,\bar{z})\right]\right\}/Z_0}{\int\mathscr{D}\varphi\exp\left[-i\int_{C_1} \frac{d\theta_1}{2\pi}\varphi(z_1,\bar{z}_1)+i\int_{C_2} \frac{d\theta_2}{2\pi}\varphi(z_2,\bar{z}_2)-S_0\right]/Z_0}\nonumber\\
&=&\frac{\left\langle\mathscr{R}\left\{\exp\left[-i\int_{C_1} \frac{d\theta_1}{2\pi}\varphi(z_1,\bar{z}_1)+i\int_{C_2}\frac{d\theta_2}{2\pi}\varphi(z_2,\bar{z}_2)\right]\left[\frac{2\pi}{L}z\partial_z\varphi(z,\bar{z})\right]\right\}\right\rangle_0}{\left\langle\mathscr{R}\left\{\exp\left[-i\int_{C_1} \frac{d\theta_1}{2\pi}\varphi(z_1,\bar{z}_1)+i\int_{C_2}\frac{d\theta_2}{2\pi}\varphi(z_2,\bar{z}_2)\right]\right\}\right\rangle_0}\nonumber\\
&\equiv&\frac{I_1}{I_2}, 
\end{eqnarray}
where $Z_0=\int\mathscr{D}\varphi\exp(-S_0)$, $\mathscr{R}$ is radial ordering and $\langle\cdots\rangle_0\equiv\int\mathscr{D}\varphi[(\cdots)\exp(-S_0)]/Z_0$ so that we could apply Wick theorem of free boson to the operator product expansion. To further use the Wick theorem, we need to write the exponential of operators into the polynomial expansions.  
\begin{eqnarray}
I_1&=&\sum_{n=0}^{+\infty}\left\langle\mathscr{R}\left\{\frac{1}{n!}\left[-i\int_{C_1} \frac{d\theta_1}{2\pi}\varphi(z_1,\bar{z}_1)+i\int_{C_2}\frac{d\theta_2}{2\pi}\varphi(z_2,\bar{z}_2)\right]^n\left[\frac{2\pi}{L}z\partial_z\varphi(z,\bar{z})\right]\right\}\right\rangle_0\nonumber\\
&=&\left\langle\mathscr{R}\left\{\left[-i\int_{C_1}\frac{d\theta_1}{2\pi}\varphi(z_1,\bar{z}_1)+i\int_{C_2}\frac{d\theta_2}{2\pi}\varphi(z_2,\bar{z}_2)\right]\frac{2\pi}{L}z\partial_z\varphi(z,\bar{z})\right\}\right\rangle_0\cdot\nonumber\\
&&\cdot\sum_{n=0}^{+\infty}\frac{n}{n!}\left\langle\mathscr{R}\left[-i\int_{C_1}\frac{d\theta_1}{2\pi}\varphi(z_1,\bar{z}_1)+i\int_{C_2}\frac{d\theta_2}{2\pi}\varphi(z_2,\bar{z}_2)\right]^{n-1}\right\rangle_0\nonumber\\
&=&I_2\cdot\left\langle\mathscr{R}\left\{\left[-i\int_{C_1}\frac{d\theta_1}{2\pi}\varphi(z_1,\bar{z}_1)+i\int_{C_2}\frac{d\theta_2}{2\pi}\varphi(z_2,\bar{z}_2)\right]\frac{2\pi}{L}z\partial_z\varphi(z,\bar{z})\right\}\right\rangle_0, 
\end{eqnarray}
where we have made the use of the Wick theorem which gives
\begin{eqnarray}
\langle\mathscr{R}(\varphi_1\cdots\varphi_k)\rangle_0=(\text{sum of all possible contractions}), 
\end{eqnarray}
where $\varphi_i$'s are linear combinition of free bosonic operators. 
Thus, after noticing $\partial_z\varphi(z,\bar{z})=\partial_z\phi(z)$ and $\mathscr{R}(\phi(z)\bar{\phi}(\bar{z}'))=0$, we obtain
\begin{eqnarray}
\label{result}
\frac{I_1}{I_2}&=&\left\langle\mathscr{R}\left\{\left[-i\int_{C_1}\frac{d\theta_1}{2\pi}\varphi(z_1,\bar{z}_1)+i\int_{C_2}\frac{d\theta_2}{2\pi}\varphi(z_2,\bar{z}_2)\right]\frac{2\pi}{L}z\partial_z\varphi(z,\bar{z})\right\}\right\rangle_0\nonumber\\
&=&\frac{2\pi}{L}z\left\langle\mathscr{R}\left\{\left[i\int_{C_2}\frac{dz_2}{i2\pi z_2}\phi(z_2)-i\int_{C_1}\frac{dz_1}{i2\pi z_1}\phi(z_1)\right]\partial_z\phi(z)\right\}\right\rangle_0\nonumber\\
&=&-i\frac{2\pi}{L}z\left[\int_{C_2}\frac{dz_2}{2\pi i}\frac{\partial_z\ln(z-z_2)}{z_2}-\int_{C_1}\frac{dz_1}{2\pi i}\frac{\partial_z\ln(z-z_1)}{z_1}\right]\nonumber\\
&=&-i\frac{2\pi}{L}z\left[\int_{|z_2|=\exp(\tau_{\text{E}0})}\frac{dz_2}{2\pi i}\frac{1}{z_2(z-z_2)}+\int_{|z_1|>|z|}\frac{dz_1}{2\pi i}\frac{1}{z_1(z_1-z)}\right]\nonumber\\
&=&-i\frac{2\pi}{L}z\left[\frac{1}{z}+\left(-\frac{1}{z}+\frac{1}{z}\right)\right]\nonumber\\
&=&-i\frac{2\pi}{L}, 
\end{eqnarray}
where we have used that $d\theta_{1,2}/(2\pi)=dz_{1,2}/(i2\pi z_{1,2})$, $\langle\mathscr{R}[\phi(z)\phi(z_{1,2})]\rangle_0=-\ln(z-z_{1,2})$, and $\partial_z\varphi(z,\bar{z})=\partial_z\phi(z)$ with $0<|z_2|<|z|<|z_1|<\infty$. Thus, Eq.~(\ref{qmw}) is obtained. 

In almost the same details, we could also derive Eq.~(\ref{qmwbar}) that $\langle\partial_{\bar{w}}\varphi(w,\bar{w})\rangle_\text{QM}=-i$ by
\begin{eqnarray}
\int_0^{2\pi}\frac{id\theta}{2\pi}\left[g(z)+g(\bar{z})\right]=2\oint_{S^1}\frac{dz}{2\pi z}g(z), 
\end{eqnarray}
for any holomorphic function $g(z)$ and $z=\exp(i\theta)$. 

Moreover, these results also imply that Eq.~(\ref{condition_1}) is true, except for that, in that case, $|z_1|=|z|$. Therefore, we apply the \emph{Cauchy principal value} regularization scheme that for any smooth contour $\gamma$ passing through finitely many poles $\{p_i\}$'s of a holomorphic function $h(z)$ and including finitely many singular points $\{q_i\}$'s of $h(z)$ in the interior of the area enclosed within $\gamma$. Then the following improper integral can be evaluated by its principal value (PV):
\begin{eqnarray}
\text{PV}\left[\int_\gamma h(z)\right]=2\pi i\sum_i\text{Res}[h(z),q_i]+\pi i\sum_i\text{Res}[h(z),p_i]. 
\end{eqnarray}
Thus, by the replacement of $|z_1|>|z|$ by $|z_1|=|z|$ in Eq.~(\ref{result}), the coefficient of $\langle0|\tilde{U}(+\infty,-\infty)|0\rangle$ in Eq.~(\ref{condition_1}) can be calculated as
\begin{eqnarray}
&&\frac{\left\langle0\left|\tilde{U}(+\infty,t_1)\left[\int_{t_1>t}\frac{dx_1}{iL}\partial_{t}\varphi(t_1,x_1)\right]\tilde{U}(t_1,-\infty)\right|0\right\rangle}{\langle0|\tilde{U}(+\infty,-\infty)|0\rangle}\nonumber\\
&=&2\left.\frac{I_1}{I_2}\right|_{|z_1|=|z|}\nonumber\\
&=&-i2z\frac{2\pi}{L}\text{PV}\left[\int_{|z_2|=\exp(\tau_{\text{E}0})}\frac{dz_2}{2\pi i}\frac{1}{z_2(z-z_2)}+\int_{|z_1|=|z|}\frac{dz_1}{2\pi i}\frac{1}{z_1(z_1-z)}\right]\nonumber\\
&=&-i2z\frac{2\pi}{L}\left[\int_{|z_2|=\exp(\tau_{\text{E}0})}\frac{dz_2}{2\pi i}\frac{1}{z_2(z-z_2)}+\text{PV}\int_{|z_1|=|z|}\frac{dz_1}{2\pi i}\frac{1}{z_1(z_1-z)}\right]\nonumber\\
&=&-i2z\frac{2\pi}{L}\left[\frac{1}{z}+\left(-\frac{1}{2}\frac{1}{z}\right)\right]\nonumber\\
&=&-i\frac{2\pi}{L}, 
\end{eqnarray}
where the factor ``$2$'' on the second line above takes into account both the holomorphic and the anti-holomorphic parts which are the same. Then we arrive at Eq.~(\ref{condition_1}). 

\subsection{Two-point correlation function}
We calculate two-point correlation function. Since there is no new technical difficulties, we directly present the results below. We define the following variables $z\equiv\exp[2\pi(\tau-ix)/L]$ and $z'\equiv\exp[2\pi(\tau'-ix')/L]$. 
\begin{itemize}
\item Density-density correlation function: 
\begin{eqnarray}
\label{density_density}
&&\langle\delta\rho(\tau,x)\delta\rho(\tau',x')\rangle\nonumber\\
&=&\frac{1}{4\pi^2}\left[\frac{zz'}{(z'-z)^2}+\text{c.c.}\right]\nonumber\\
&\rightarrow&L^2\left\{\frac{1}{\left[(\tau-\tau')-i(x-x')\right]^2}+\frac{1}{\left[(\tau-\tau')+i(x-x')\right]^2}\right\}, 
\end{eqnarray}
as $|(\tau-\tau')+i(x-x')|\ll L$: the thermodynamical limit. We can see that the density-density correlation does not see the pulse, which reflects that the translational invariance of the system is always present. 
\item Density-current correlation function: 
\begin{eqnarray}
\langle\delta\rho(\tau,x)J(\tau',x')\rangle=0. 
\end{eqnarray}
\item Current-current correlation function: $\left(v_\text{F}\equiv1\right)$
\begin{eqnarray}
&&\langle J(\tau,x)J(\tau',x')\rangle\nonumber\\
&=&\frac{2\Theta(t-t_0)}{L}\frac{2\Theta(t'-t_0)}{L}+\frac{1}{2\pi^2}\text{Re}\frac{zz'}{(z'-z)^2}\nonumber\\
&=&\langle J(\tau,x)\rangle\langle J(\tau',x')\rangle+\langle\delta\rho(\tau,x)\delta\rho(\tau',x')\rangle, 
\end{eqnarray}
where the second term, of the same value as Eq.~(\ref{density_density}), is the correlation in the absence of pulse while the first term implies that the current operator gains a nonzero expectation value after the pulse. 
\end{itemize}
\subsection{Tomonaga-Luttinger liquid}
\label{TL_liquid}
Let us consider the following ``spinless'' Tomonaga-Luttinger model (bosonization of some interacting fermionic model) in Euclidean signature: 
\begin{eqnarray}
S^{0}_\text{T-L}&=&\frac{1}{8\pi K}\int d\tau_\text{E}dx\left[(\partial_{\tau_\text{E}}\varphi)^2+(\partial_x\varphi)^2\right], 
\end{eqnarray}
where $\tau_\text{E}\equiv ivt$ with $K$ and $v$ normalizing factors for Luttinger parameter $1/4\pi$ and Fermi velocity $v_\text{F}$, respectively, due to the interaction of the corresponding fermionic model. To introduce the background gauge field coupled with $\varphi$, we could use
\begin{eqnarray}
[\varphi(t,x),\theta(t,x')]=i2\pi\Theta(x'-x), 
\end{eqnarray}
where $\theta$ is the dual field to $\varphi$ and it is physically the $U(1)$ electromagnetic phase of Dirac spinor. Then, after taking the derivatives of $x'$ and $x$, respectively, above, we obtain that
\begin{eqnarray}
&&\left[\varphi(t,x),\frac{1}{2\pi}\partial_{x'}\theta(t,x')\right]=i\delta(x-x')=[\varphi(t,x),\Pi_\varphi(t,x')]\nonumber\\
&&\left[\frac{1}{2\pi}\partial_{x}\varphi(t,x),\theta(t,x')\right]=-i\delta(x-x')=[\Pi_\theta(t,x),\theta(t,x')]
\end{eqnarray}
These canonical relations implies the first duality mapping: 
\begin{eqnarray}
\label{dual_1}
\frac{1}{4\pi Kv}\partial_t\varphi\leftrightarrow\frac{1}{2\pi}\partial_x\theta. 
\end{eqnarray}
Thus we can derive the $S^0_\text{T-L}$ in terms of $\theta$ field as
\begin{eqnarray}
S^0_\text{T-L}=\frac{K}{2\pi}\int d\tau_\text{E}dx\left[(\partial_{\tau_\text{E}}\theta)^2+(\partial_x\theta)^2\right], 
\end{eqnarray}
which gives us the second duality mapping as
\begin{eqnarray}
\label{dual_2}
\frac{K}{\pi v}\partial_t\theta\leftrightarrow\frac{1}{2\pi}\partial_x\varphi. 
\end{eqnarray}
The $\theta$-representation of $S^0_\text{T-L}$ is useful to apply the minimal coupling by $\vec{\partial}\theta\rightarrow\vec{\partial}\theta+e\vec{A}$. Then we apply the duality mappings of Eqs.~(\ref{dual_1},\ref{dual_2}) and restore the gauge invariance and anomaly-matching. Therefore, the $\varphi$-representation of action follows as, e.g. on a torus $T^2$, 
\begin{eqnarray}
S_\text{T-L}[A]=S^0_\text{T-L}-i\frac{e}{2\pi}\int_{T^2}\varphi dA-i\frac{\epsilon^{\mu\nu}}{2\pi}\left(\int_{\text{cycle}_\mu}d\varphi\right)\left(\int_{\text{cycle}_\nu}eA\right)
\end{eqnarray}
with the radius $R=1$. 
\begin{eqnarray}
\label{density_1}
\delta\rho\leftrightarrow-\frac{1}{2\pi}\partial_x\varphi&=&\frac{i}{2\pi}(\partial_w-\partial_{\bar{w}})\varphi(w,\bar{w})\nonumber\\
J\leftrightarrow+\frac{1}{2\pi}\partial_t\varphi&=&i\frac{v}{2\pi}(\partial_w+\partial_{\bar{w}})\varphi(w,\bar{w}), 
\end{eqnarray}
where again $w=\tau_\text{E}-ix$, $\bar{w}=\tau_\text{E}+ix$, but $\tau_\text{E}= ivt$ as defined above. 
\subsection{Spatially uniform electromagnetic pulse: $eF_{01}=2\pi\delta(t-t_0)/L$}
Then we consider the geometry of an infinite-long cylinder with circumference $L$. As before, we do the conformal mapping: $z=\exp(2\pi w/L)$. Then, 
\begin{eqnarray}
J&=&i\frac{v}{2\pi}\left[(\partial_wz)\partial_z+(\partial_{\bar{w}}\bar{z})\partial_{\bar{z}}\right]\varphi(z,\bar{z})\nonumber\\
&=&i\frac{v}{L}\left(z\partial_z+\bar{z}\partial_{\bar{z}}\right)\varphi(z,\bar{z}), 
\end{eqnarray}
and, similarly, 
\begin{eqnarray}
\delta\rho=\frac{i}{L}\left(z\partial_z-\bar{z}\partial_{\bar{z}}\right)\varphi(z,\bar{z}). 
\end{eqnarray}
To evaluate the correlation functions, we still use the following Minkowskian action: 
\begin{eqnarray}
\tilde{S}_\text{T-L}(t_1)&=&S^0_\text{T-L}-\int_{t_1>t}\frac{dx_1}{L}\varphi(t_1,x_1)+\int_{t_2=t_0}\frac{dx_2}{L}\varphi(t_2,x_2). 
\end{eqnarray}
The effect of $K\neq1$ is the modified correlation function: 
\begin{eqnarray}
\langle\varphi(z,\bar{z})\varphi(z',\bar{z}')\rangle^\text{T-L}_0=-K\ln[(z-z')(\bar{z}-\bar{z}')]. 
\end{eqnarray}
Then we could calculate that
\begin{eqnarray}
\label{phase_1}
\langle0|\tilde{U}_\text{T-L}(+\infty,-\infty)|0\rangle=\exp\left[-ivK\frac{2\pi}{L}(t_1-t_0)\right], 
\end{eqnarray}
which means $\Delta E_\text{T-L}=2\pi vK/L$ the excitation energy and it exactly coincides with that of a $\theta$-vortex quanta excitation. 

Similarly, 
\begin{eqnarray}
\langle\delta\rho(t,x)\rangle&=&0,\nonumber\\
\langle J(t,x)\rangle&=&\frac{2K\Theta(t-t_0)}{L}v. 
\end{eqnarray}

\subsection{Lattice momentum}
\label{momentum_pulse_sec}
To calculate the lattice momentum $P_\text{lattice}a$ with $a$ the lattice constant, we need its field-theoretical representation in terms of $\varphi$. Since
\begin{eqnarray}
\exp(iP_\text{lattice}a)\varphi\exp(-iP_\text{lattice}a)=\varphi+2\pi\nu, 
\end{eqnarray}
from which we obtain that $Q_\text{trans}$ is exactly the Noether charge related to $\varphi\rightarrow\varphi+2\pi\nu$: 
\begin{eqnarray}
\label{momentum_pump_0}
P_\text{lattice}a&=&\int dx\,2\pi\nu\Pi_\varphi\mod 2\pi\nonumber\\
&=&\int dx\frac{\pi\nu}{2\pi Kv}\partial_t\varphi\mod 2\pi\nonumber\\
&=&\int dx\frac{\pi\nu}{Kv}J\mod 2\pi. 
\end{eqnarray}

Thus
\begin{eqnarray}
\label{momentum_pump_2}
\langle P_\text{lattice}a\rangle=\Theta(t-t_0){2\pi\nu}\mod2\pi, 
\end{eqnarray}
which needs no renormalization due to its topological nature after the duality relation in Eq.~(\ref{dual_1}) being considered: 
\begin{eqnarray}
P_\text{lattice}a&=&\int dx\frac{k_\text{F}}{2\pi Kv}\partial_t\varphi\mod2\pi\nonumber\\
&=&\int dxk_\text{F}\frac{\partial_x\theta}{\pi}\mod2\pi\nonumber\\
&=&2\pi\nu N_\theta\mod2\pi, 
\end{eqnarray}
where $N_\theta$ is the winding number of $\theta$ field thereby necessarily being integer-valued. Comparing with Eq.~(\ref{momentum_pump_2}) and $\Delta E_\text{T-L}=2\pi vK/L$, we confirm that the unit external pulse indeed exactly excites a unit $\theta$-vortex rather than an oscillating mode. 
\end{widetext}


\sloppy
%

\end{document}